\documentclass[aps,prb,twocolumn,groupedaddress,floatfix,showpacs]{revtex4-1}
\usepackage[T1]{fontenc}
\usepackage{graphicx}
\usepackage{amsmath}
\usepackage{color}
\newcommand{\parl}{\parallel}
\begin{document}

%\preprint{}

\title{Rydberg Magnetoexcitons in Cu$_2$O Quantum Wells}


\author{David Ziemkiewicz}
\email{david.ziemkiewicz@utp.edu.pl}
\author{Gerard Czajkowski}
\author{Karol Karpi\'{n}ski}
\author{Sylwia
Zieli\'{n}ska-Raczy\'{n}ska}
%\thanks{}
 \affiliation{Institute of
Mathematics and Physics, UTP University of Science and Technology,
\\Aleje Prof. S. Kaliskiego 7, 85-789 Bydgoszcz, Poland.}


\date{\today}

\definecolor{green}{rgb}{0,0.8,0}

\begin{abstract}
We present  theoretical approach that allows for calculation of optical functions for
Cu$_2$O Quantum Well (QW) with Rydberg excitons in an external magnetic field of an arbitrary field strength. Both Faraday and Voigt configurations are considered, in the energetic region of p-excitons.  We use the real density matrix approach and an effective e-h potential, which enable to derive analytical expressions for the QW magneto-optical functions. For both configurations, all three field regimes: weak, intermediate, and high field, are considered and treated separately. With the help of the developed approximeted method we are able to estimate the limits between the field regimes.
%Choosing the susceptibility, we performed numerical calculations appropriate to Cu$_2$O QWs.
The obtained theoretical magneto-absorption spectra show a good agreement with available experimental data.
\end{abstract}
\pacs{71.35.-y,78.20.-e,78.40.-q}
\maketitle

\section{Introduction}
The discovery of Rydberg excitons (REs) in cuprous oxide, first
observed by Kazimierczuk \emph{et al}
\cite{Kazimierczuk} initiated a large number of studies on their
spectroscopic and optical properties, see the review paper
\cite{AssmannBayer_2020}, where the extensive list of
references can be found. A lot of attention has been devoted to the
interaction of REs with an external electric and/or
magnetic field (Stark and Zeeman effects)
Refs.\cite{Rommel}$^-$\cite{Zielinska.PRB.2016.c}
%\cite{Thewes,Schoene,FS95,Magnetoexcitons_2019,Zielinska.PRB,Zielinska.PRB.2016.b,Zielinska.PRB.2016.c}
and these phenomena
have been studied, both experimentally and
theoretically  in bulk crystals or in plane-parallel slabs with
dimensions much greater than the incident wave length and the
effective Bohr radius.  The exciton Rydberg energy in Cu$_2$O of about 90 meV is lower (regarding modulus) by the order of magnitude comparing with typical semiconductors (e.g., 4.2 meV in the prototypical semiconductor GaAs
with n=3 as highest observed state). This reduction makes REs sensitive to external fields. % and enables one to reach  the strong field regime, where the characteristic interaction energies with a magnetic field dominante the Coulomb interaction.

In low dimensional systems, due to confinement effects, the excitonic states have larger energy
and oscillator strengths as compared to the bulk. This is also true for systems with
REs; the states with large main quantum number  gain additional energy therefore one can expect that depending on the type of confinement, new states can
appear, originating from the overlapping of confinement states with the Coulomb states and eventually resulting from Zeeman splitting  in an external magnetic field.
Recently Cu$_2$O  based nanostructures with REs have
 awoked an interest of several groups
\cite{Ziemkiewicz_PRB_2020}$^,$\cite{Naka}$^-$\cite{Konzelmann},
because one can expects  interesting quantum effects arising from competition between the  geometric confinement, excitons motion  and  their  interaction with additional external fields.
The interest is also motivated by virtual benefits, because
  quantum-confined structures with REs might be of use in practice for costructing new class of optoelectronic aparatus. From practical point of view one has to mention that devices such as lasers, photodetectors, modulators, and switches based on quantum wells,
  turned out to be more faster then conventional electrical components, therefore they
 are desirable for technology and telecomunicatiom. It might be interesting to consider  a possibility of an additional manipulation on demand by an external magnetic field applied to quantum wells with REs.

Similarly to  the bulk case, an application of external
fields to nanostructures changes the spectra of REs \cite{Ziemkiewicz_PRB_2020}.  Since the electro- and magneto-optical
properties of typical, the most studied case of GaAs based nanostructures have been explored  for
decades, the discussion of the Cu$_2$O, when multiple Rydberg
exciton states must be taken into account, is already at the
beginning. Here we will consider the effect of an external magnetic field on  a quantum well with Rydberg excitons.  The effects of a geometric confinement is superimposed on
  REs interaction with external fields, it manifests in intrinsic difference of magneto-optical spectra in terms of state energies, which in turn depend on a field orientation. Inspired by the recent development in the area of REs, we aim to analyze
the magneto-optical properties of Cu$_2$O based quantum well (QW) in two diffrent field orientation, namely the Faraday and Voight configurations.
%In the first situation the magnetic field is applied along the optical axis, while in the second case the field is oriented normal to optical axis.
Both cases have been investigated for bulk Cu$_2$O crystals with RE for weak magnetic field(up to 4 T) experimentally and the numerical excitonic spectra were shown \cite{Rommel}.

We will use the real density matrix approach (RDMA) to calculate
the optical functions of a single Cu$_2$O QW with REs. This approach turned out to be
 successful in describing the optical properties of Cu$_2$O
bulk crystals, including effects of external fields
Ref.\cite{Magnetoexcitons_2019}$^,$
\cite{Zielinska.PRB.2016.c}$^,$\cite{Zielinska.PRB.2016a}. As it was
shown in our recent paper \cite{Ziemkiewicz_PRB_2020}  it is
possible  to extend the RDMA method for low dimensional systems.
When describing the magneto-optical properties of the systems with
excitons, one is confronted with well known difficulties. The
exciton, being an analogy of a hydrogen atom, is created and
maintained by a Coulomb attraction, having a spherical symmetry.
On the other hand, in the case of a quantum well a magnetic field
and the confinement potentials  have a cylindrical symmetry. These
geometrical discerepancies  rule out an analytical solution of the
proper Schr\"{o}dinger equation for the problem. To circumvent
such obstacles  we have to use, as in the bulk case, various
approximations, which depend on the relation between the exciton
binding energy and the magnetic field energy. When the excitonic
energies are larger than the magnetic field energies (the Landau
states energies), we use the so-called weak field approximation.
In the opposite case, when the Landau states energies  are greater
than the excitonic state energies we have to consider a high field
approximation. Between these two regions one has to consider the
intermediate field case, when both, the excitonic energies and the
Landau state energies, are comparable. Moreover, each magnetic
field regimes requires a different theoretical approach and there
is strong need for a versatile estimation how to distinct the
regime of the magnetic field; we propose the method, which allow
to discern these regimes. When concerning the magneto-optical
properties of excitons in a QW, one has to account for effects
related to the direction of the applied magnetic field. One
distinguishes between the Faraday configuration, when the magnetic
field is directed along the growth axiss (the $z$-axis,
perpendicular to the planes of the QW), and the Voigt
configuration, when the magnetic field is perpendicular to the
$z$-axis and parallel to the planes of the QW.  We will show
below, that all the above mentioned effects can be described
within the RDMA.

The paper is organized as follows. In Sec.
\ref{basic_equations} we recall the basic equations of RDMA,
adapted to the case of QWs, when external fields are applied. In
the next three sections we explicitly derived the formulas for
magneto-suscptibility for Cu$_2$O QWs when the external magnetic
field is applied in the Faraday configuration. We separately
discussed the cases of a week field (Sec. \ref{Faraday_weak}),
high field (Sec. \ref{Faraday_high}), and the intermediate
magnetic field (Sec. \ref{Farady_intermediate}). Then we will also
consider three different regimes of the magnetic field strength in
the case of the Voigt configuration (Sections
\ref{Voigt_weak}-\ref{Voigt_intermediate}). Sec. \ref{Results}
contains illustrative numerical results and the description of a simple but effective method, which allows for estimation of the distintion between magnetic field regimes while a summary and
conclusions of our paper are presented in Sec. \ref{conclusions}. Four Appendices contain the details of analytical calculations.

\section{Basic equations}\label{basic_equations}
We will use the real density matrix
 approach, applied to single quantum well with   Rydberg
 states, similary as it was done for low dimension structures  in Ref. \cite{Ziemkiewicz_PRB_2020}
  In this approach the optical properties are described by an
equation for the coherent amplitudes $Y_{12}$ of the electron-hole
pair of coordinates ${\bf r}_1={\bf r}_h$ and ${\bf r}_2={\bf
r}_e$ which for a pair of conduction and valence bands
\begin{equation}\label{Ylinear}
-i(\hbar \partial_t + \Gamma) Y_{12}+H_{eh}Y_{12} ={\bf M}{\bf
E},
\end{equation}
\noindent where ${\bf E}$ is the electric field, $\Gamma$ is a
phenomenological damping coefficient, ${\bf M}(\textbf{r})$ is a
smeared-out transition dipole density which depends on the
coherence radius $r_0 = \left[(2\mu/\hbar^2)E_g\right]^{-1/2}$ and
the $E_g$ is the fundamental gap; $\mu$ is reduced effective mass
of the electron-hole pair and \textbf{r} is the relative
electron-hole distance. \cite{Zielinska.PRB.2016a} Specific forms
of {\bf M}(\textbf{r}) will be defined in subsequent sections.

RDMA, adopted for semiconductors by Stahl, Balslev, and others
\cite{Stahl} is a mesoscopic approach,
which in the lowest order neglects all effects from the
multiband semiconductor structure, so that the exciton Hamiltonian
becomes identical to the two-band effective mass Hamiltonian
$H_{eh}$, which in the case when external fields are applied,
includes the electron and hole kinetic energy, the electron-hole
interaction potential, the terms related to the external fields,
and the confinement potentials.\cite{Hecktotter_2018} In
consequence, the Hamiltonian $H_{eh}$ is given by
\begin{eqnarray}\label{BFmagnetichamiltonian1}
&&\phantom{nucl}H=E_{g}+\frac{1}{2m_e} \left({\bf p}_e-e
\frac{{\bf r}_e \times {\bf B}}{2}\right)^2\nonumber\\
&& + \frac{1}{2m_{h}} \left({\bf p}_h + e \frac{{\bf r}_h \times
{\bf B}}{2}\right)_z^2 \nonumber\\ & &+ \frac{1}{2m_{h}} \left(
{\bf p}_h+e \frac{{\bf r}_h \times {\bf B}}{2}\right)_\parl^2
+e{\bf F}\cdot({\bf r}_e-{\bf r}_h) \nonumber\\
&&+ V_{\rm conf}({\bf r}_e,{\bf r}_h) -
\frac{e^2}{4\pi\epsilon_0\epsilon_b \vert{\bf r}_e - {\bf
r}_h\vert} ,
\end{eqnarray}
 \noindent
 {\bf B} is the magnetic field
vector, ${\bf F}$ the electric field vector,  $V_{\rm conf}$ are
the surface potentials for electrons and holes, $m_{hz}, m_{h
\parl}$ are the components of the hole effective mass tensor, and the
electron mass is assumed to be isotropic.  The total polarization
of the
medium is related to the coherent amplitude by
\begin{equation}\label{Polar}
{\bf P}({\bf R})=2 \hbox{Re}\int d^3{r}\,{\bf
M}({\bf r}) Y({\bf R},{\bf r})
\end{equation}
where $\textbf{R}$ is the center-of-mass coordinate. This, in turn, is used in Maxwell's field equation
\begin{equation}
c^2\nabla^2 {\bf E(R)} - \epsilon_b \ddot{\bf E} = \frac{1}{\epsilon_0}{\bf \ddot{P}(R)}.
\end{equation}
The excitonic susceptibility $\chi$ is then given by
\begin{equation}\label{susceptibility_Fourier}
{\bf P(\omega,\textbf{k})} =\epsilon_0\chi(\omega,\textbf{k}){\bf E}(\omega,\textbf{k})
\end{equation}
where $\omega$ is the frequency of the incident field and the absorption coefficient can be calculated from
\begin{eqnarray}\label{abscoeff1}
{\bf \alpha}=2\frac{\hbar\omega}{\hbar
c}\hbox{Im}\,\sqrt{\epsilon_b+\chi},
\end{eqnarray}
where $\epsilon_b$ is the background dielectric constant. The
detailed form of the Hamiltonian  for  both Faraday and Voigt configurations will be applied in the following sections.

\section{The Faraday configuration}When the magnetic field \textbf{B} is
applied to a QW in the growth direction, which we identify with
the $z$-axis, we deal with the Faraday configuration.
\subsection{Weak field
limit}\label{Faraday_weak}  In this
configuration we will consider the optical response of the QW with
thickness $L$ to a normally incident electromagnetic wave. The QW
is located in the $x-y$ plane, with the surfaces at $z=\pm L/2$.
We can separate the motion in the $z$-direction (where the
particles are treated separately) from the in-plane motion where
we use the relative- and exciton center-of-mass coordinates.
In the case of $F=0$ we transform the Hamiltonian
(\ref{BFmagnetichamiltonian1}) into the form

\begin{eqnarray}\label{SL3Hamiltonian}
H&=&H_0+\frac{P_z^2}{2M_z} + \frac{{\bf P}_\parl^2}{2M_\parl} +
\frac{1}{8}\mu
\omega_c^2 r_\parl^2 + \frac{e}{2\mu'}B {\mathcal L}_z   \nonumber\\
& & -\frac{e}{M_\parl} {\bf P}_\parl\cdot \left( {\bf
r}_\parl\times{\bf B} \right)+V_{\rm conf}({\bf r}_e,{\bf r}_h),
\end{eqnarray}
where $\omega_c={eB}/{\mu_\parl}$ is the cyclotron frequency,
\noindent the reduced mass $\mu'$ is defined as
\begin{equation}\label{reducedmuprim}
 \frac{1}{\mu'}=\frac{1}{m_e}-\frac{1}{m_{h}},
\end{equation}
\noindent and $H_0$ is the two-band Hamiltonian for the relative
electron-hole motion, as  used in the papers.
\cite{Zielinska.PRB}$^,$ \cite{Zielinska.PRB.2016.b} The operator
${\mathcal L}_z$ is the \emph{z}-component of the angular momentum
operator.

We assume a parabolic confinement in the $z$-direction,
\begin{equation}\label{confinement}
V_{conf}=\frac{1}{2}m_e\omega_{ez}^2z_e^2+\frac{1}{2}m_h\omega_{hz}^2z_h^2,\end{equation}
and using the notation
\begin{eqnarray}
&&H^{(1D)}_{m,\omega}(z)=\frac{p_z^2}{2m}+\frac{1}{2}m\omega^2z^2,
\end{eqnarray}
the QW Hamiltonian can be written in the form
\begin{eqnarray}\label{HmagneticQW}
&&H_{QW}=E_g+H_{m_{e},\omega_{ez}}^{(1D)}(z_e)+H_{m_{h},\omega_{hz}}^{(1D)}(z_h)\nonumber\\
&&-\frac{\hbar^2}{2M_{z}}\partial^2_Z-
 \frac{\hbar^2}{2M_{\parl }}\hbox{\boldmath$\nabla$}^{(2D)2}_{R_{\parl}}
- \frac{\hbar^2}{2\mu}
\hbox{\boldmath$\nabla$}^{(2D)2}_{r}\nonumber
\\
&&-\frac{\mu}{\mu'}{ i}\gamma R^*\partial_{\phi}
+\frac{R^*}{4a^{*2}}\gamma^2\,r^2+V_{eh},
\end{eqnarray}
\noindent where
 $V_{eh}$ is the electron-hole Coulomb interaction potential, $a^*$  is the exciton Bohr radius and $R^*$ the exciton Rydberg
 energy,
 $\hbox{\boldmath$\nabla$}^{(2D)2}_{R_{\parl}},\hbox{\boldmath$\nabla$}^{(2D)2}_{r}$
 denote 2-dimensional nabla operators, and $r=\sqrt{x^2+y^2}$.
 The dimensionless strength of the magnetic field $\gamma$ is defined as
\begin{equation} \label{gamma}
 \gamma=\hbar\omega_c/2R^*,
 \end{equation}

 In the weak magnetic field limit excitons play a dominant role in
 determining the optical response, the magnetic field can be
 treated as a perturbation\cite{Magnetoexcitons_2019} and
 we use
 the 2-dimensional Coulomb potential
 \begin{equation}
 V_{eh}=-\frac{e^2}{4\pi\epsilon_0\epsilon_b r}.\end{equation}
 With respect to the above assumptions, the l.h.s. operator in Eq.
(\ref{HmagneticQW}) includes two one dimensional harmonic
oscillator Hamiltonians and the 2-dimensional Coulomb Hamiltonian
\begin{equation}\label{2-dim_Coulomb}
H^{(2D)}_{Coulomb}=- \frac{\hbar^2}{2\mu}
\hbox{\boldmath$\nabla$}^{(2D)2}_{r}-\frac{e^2}{4\pi\epsilon_0\epsilon_b
r}.
\end{equation}
We also neglect the terms related to the center-of-mass motion.
Therefore the solution for the amplitude $Y$ can be expressed in
terms of eigenfunctions of the mentioned Hamiltonians
\begin{eqnarray}\label{expansion}
&&Y_{jmN_eN_h}=\\
&&=\sum\limits_{N_e,N_h,j,m}
c_{jmN_eN_h}\psi^{(1D)}_{\alpha_{ez,N_e}}(z_e)\psi^{(1D)}_{\alpha_{hz,N_h}}(z_h)
\psi_{jm}(r,\phi),\nonumber
\end{eqnarray}
where  $\psi^{(1D)}_{\alpha_{z},N}(z)$ ($N_e$,$N_h$=0,1,...) are
the quantum oscillator eigenfunctions for electron and hole,
respectively.
\begin{eqnarray}\label{eigenf1doscillator}
&&
 \psi^{(1D)}_{\alpha_{z},N_{e,h}}(z)=
 \pi^{-1/4}\sqrt{\frac{\alpha_{z}}{2^N_{e,h} N_{e,h}!}} H_N(\alpha_{z}z)
e^{-\frac{\alpha_{z}^2}{2}z^2}, \nonumber\\
&&
 \alpha_{z} = \sqrt{\frac{m_{e,h} \omega_{z}}{\hbar}},
\end{eqnarray}
 $H_N(x)$ are Hermite polynomials $(N_{e,h}=0,1,\ldots)$, $m_{e,h}$ are the electron (hole) effective
 masses, and $\psi_{jm}(\rho,\phi)$ are the eigenfunctions of the
 2-dimensional Hamiltonian (\ref{2-dim_Coulomb})
 \begin{eqnarray}\label{2_Dim_Eigen}
 &&\psi_{jm}(\rho,\phi)=R_{jm}(\rho)\frac{e^{im\phi}}{\sqrt{2\pi}},\nonumber\\
&&R_{jm}=A_{jm}e^{-2\lambda\rho}(4\lambda\rho)^{\vert
m\vert}L^{\vert 2
m\vert}_j(4\lambda\rho),\\
&&\lambda=\frac{1}{1+2(j+\vert m\vert)},\nonumber\\
&&A_{jm}=\frac{4}{(2j+2\vert
m\vert+1)^{3/2}}\left[\frac{j!}{(j+2\vert
m\vert)!}\right]^{1/2},\nonumber
\end{eqnarray}
where $L^\alpha_n(x)$ are the Laguerre polynomials, for which we
use the definition
\begin{displaymath}
L^\alpha_n(x)={n+\alpha\choose
n}M(-n,\alpha+1;x),\end{displaymath} with  the Kummer function
$M(a,b,z)$ (the confluent hypergeometric function),\cite{Abramowitz},
$\rho=r/a^*$ is the scaled space variable. Here
we use the transition dipole density in the form
\cite{Magnetoexcitons_2019}
\begin{equation}\label{dipoledensity}
M(\hbox{\boldmath$\rho$},z_e,z_h)=\frac{M_0}{2\rho_0^3}\rho\,e^{-\rho/\rho_0}\frac{e^{i\phi}}{\sqrt{2\pi}}\delta(z_e-z_h),
\end{equation}
with the integrated strength $M_0$ and the coherence radius
$\rho_0=r_0/a^*$, where $r_0=\sqrt{\frac{\hbar^2}{2\mu E_g}}$. The coefficient $M_0$ and the coherence radius
$\rho_0$ are connected through the longitudinal-transversal energy
$\Delta_{LT}$ \cite{Zielinska.PRB.2016a}
\begin{equation}
(M_0\rho_0)^2=\frac{4}{3}\frac{\hbar^2}{2\mu}\epsilon_0\epsilon_ba^*\frac{\Delta_{LT}}{R^*}\,e^{-4\rho_0}.
\end{equation}
The calculation of the QW susceptibility, from which other optical
functions can be determined, consists of several steps. First, we
assume that the incident electromagnetic wave is linearly
polarized with the electric vector \textbf{E} with a component
in the direction $\alpha$ and an amplitude $\mathcal E$; the
dipole density vector \textbf{M} has a component $M$ in the form
(\ref{dipoledensity}) in the same direction. Than, with the help od Eqs
(\ref{Polar})and (\ref{susceptibility_Fourier}), applying the long wave
approximation, we calculate the mean QW susceptibility from  the
formula
\begin{eqnarray}\label{def_mean_susceptibility}
&&\chi=\frac{2}{\epsilon_0{\mathcal
E}}\frac{1}{L}\\
&&\times\int\limits_{-L/2}^{L/2}dz_edz_h\,d^2
\hbox{\boldmath$\rho$}M(\hbox{\boldmath$\rho$},z_e,z_h)Y(\hbox{\boldmath$\rho$},
z_e,z_h).\nonumber
\end{eqnarray}
The first step to
calculate  $\chi$ is to determine the exciton amplitude $Y$.
Finally we use  Eq. (\ref{Ylinear}) with  the Hamiltonian  given by Eq.
(\ref{HmagneticQW}). Inserting the expansion
(\ref{expansion}) into Eq. (\ref{Ylinear}) and making use of the
dipole density in the form (\ref{dipoledensity}), one obtains a
set of linear algebraic equations for the expansion coefficients
$c_{jmN_eN_h}$
\begin{eqnarray}\label{system_weak_F}
&&\sum_{\ell=0}^{j_{max}}a_{j\ell mN_eN_h}c_{
\ell mN_eN_h}= b_{j1}\delta_{N_eN_h}{\mathcal E},\nonumber\\
&&a_{j\ell mN_eN_h}=\delta_{j\ell}\kappa^2_{jmN_eN_h}+V_{j\ell m},\nonumber\\
&&\kappa^2_{jmN_eN_h}=\frac{1}{R^*}\Biggl(E_g-\hbar\omega-i{\mit\Gamma}+\varepsilon_{jm}R^*
+W_{eN_e}\nonumber\\
&&+W_{hN_h}+\frac{\mu}{\mu'}m\gamma\,R^*\Biggr),\nonumber\\
&&\epsilon_{jm}=-4\lambda_{jm}^2,\nonumber\\
&&\lambda_{jm}=\frac{1}{2j+2\vert m\vert +1},\\
&&b_{j\vert m\vert
N_eN_h}=b_{j1N_eN_h}\nonumber\\
&&=\sqrt{\frac{(j+1)(j+2)}{(j+3/2)^5}}(1+2\rho_0\lambda_{j1})^{-4}
F\left(-j,4;3;\frac{1}{s}\right)\nonumber\\
&&s=\frac{1+2\rho_0\lambda_{jm}}{4\rho_0\lambda_{jm}},\nonumber\\
&&j,\ell=0,1,2,\ldots,j_{max},\quad m=\pm 1,\quad
N_e,N_h=0,1,2,\ldots,\nonumber
\end{eqnarray}
where $F(\alpha,\beta;\gamma;z)$ is a hypergeometric series.
$V_{j\ell m}$ are matrix elements
\begin{equation}\label{matrix_elem_Farad_week}
V_{j\ell  m}=\frac{1}{4}\gamma^2\langle R_{jm}(\rho)\vert
\rho^2\vert R_{\ell m}(\rho)\rangle,
\end{equation}
and their detailed form is given by Eq.
(\ref{M_Elem_weak}) in Appendix \ref{Appendix A}. The $z$-confinement energies $W_{eN_e},
W_{hN_h}$ and the parameters $\alpha$ are defined as follows
\begin{eqnarray*}
&&\alpha_{e}=\sqrt{\frac{m_e}{\mu}}\sqrt{\frac{W_{e0}}{R^*}},\\
&&\alpha_{h}=\sqrt{\frac{m_h}{\mu}}\sqrt{\frac{W_{h0}}{R^*}},\nonumber\\
&&p=\frac{1}{2}\left(\alpha_{ez}^2+\alpha_{hz}^2\right),\nonumber\\
&&W_{e0}=\left(\frac{\pi a_e^*}{L}\right)^2R^*_e,\nonumber\\
&&W_{h0}=\left(\frac{\pi a_h^*}{L}\right)^2R^*_h,\nonumber\\
&&W_{e1}=3\,W_{e0},\\
&&W_{h1}=3\,W_{h0}.\\
\end{eqnarray*}
For the case  $\alpha_e=\alpha_h=\alpha^F=\pi/L$ the specific values of these parameters are
\begin{eqnarray*}
&&p=1,\nonumber\\
&&W_{e0}=\left(\frac{\pi a_e^*}{L}\right)^2R^*_e,\nonumber\\
&&W_{h0}=\left(\frac{\pi a_h^*}{L}\right)^2R^*_h,\nonumber\\
&&\frac{W_{e0}+W_{h0}}{R^*}=\left(\frac{\pi
a^*}{L}\right)^2=:\frac{W_{eh0}}{R^*},\\
&&\frac{W_{e1}+W_{h1}}{R^*}=\frac{3\,W_{eh0}}{R^*},\\
&&\frac{W_{eN}+W_{hN}}{R^*}=\frac{(2N+1)\,W_{eh0}}{R^*}.
\end{eqnarray*}
With the above definitions, taking $N_e=N_h=N$ with
computed $c$ coefficients, we use them in the expansion
(\ref{expansion}), which is in turn inserted into the Eq.
(\ref{def_mean_susceptibility}), from which  we calculate the mean
QW magneto-susceptibility for the Faraday
configuration
\begin{eqnarray}\label{chiF_weak}
&&\chi^{F}(\omega)=48\epsilon_b\frac{\Delta_{LT}}{R^*}\left(\frac{a^*}{L}\right)\nonumber\\
&&\times\sum_{j=0}^{\mathcal N}b_{j1}\left[\langle
\Psi_{00}\rangle_L\left( c_{j100}+c_{j-100}\right)+
\langle\Psi_{11}\rangle_L\left(c_{j111}+c_{j-111}\right)\right.\nonumber\\
&&\left.+\ldots
\langle\Psi_{NN}\rangle_L\left(c_{j1NN}+c_{j-1NN}\right)\right]\\
 &&\langle
\Psi_{NN}\rangle_L=\frac{1}{2^N\,N!}\frac{2}{\sqrt{\pi}}\int\limits_0^{\alpha
L/2}e^{-t^2}H_N^2(t)dt.\nonumber\end{eqnarray}

\subsection{High field
limit}\label{Faraday_high}
In the high field limit the magnetic energy contributions to the Hamiltonian are much greater then the Coulomb one and the
energies of Landau states are larger than the absolute value of the
lowest exciton state. Therefore we seek solutions for the exciton
amplitude $Y$ in terms of the eigenfunctions of the
"kinetic+magnetic+confinement\," part of the Hamiltonian
(\ref{HmagneticQW}).
\begin{equation}\label{expansionHF}
Y=\sum\limits_{nmN_eN_h}c_{nmN_eN_h}R_{nm}(\rho)\frac{e^{im\phi}}{\sqrt{2\pi}}\Psi_{N_eN_h}(z_e,z_h),
\end{equation}
where
\begin{eqnarray}\label{magnetic_eigen_f}&& R_{n m}(\rho)=\\
&&=\sqrt{\gamma}\sqrt{\frac{n!}{(n+\vert
m\vert)!}}\left(\frac{\gamma\rho^2}{2}\right)^{\vert
m\vert/2}e^{-\gamma\rho^2/4}L_{n}^{\vert
m\vert}\left(\frac{\gamma\rho^2}{2}\right),\nonumber\end{eqnarray}
  $n= 0,1,\ldots$ and $m$ depict Landau states,
 $L_n^{\vert m\vert}$ are Laguerre polynomials. Similar as in the case of weak magnetic fields, we insert the
 expansion (\ref{expansionHF}) into the Eq.(\ref{Ylinear}) with
 an appropriate form of the Hamiltonian $H_{eh}$, to obtain the expansion
 coefficients $c$, which are calculated from the set of linear equations
 \begin{eqnarray}\label{coeffHF}
&&\sum_{n mN_eN_h}a_{n\ell mN_eN_h}c_{n mN_eN_h}=d_{\ell m}\delta_{N_eN_h},\nonumber\\
&&a_{n\ell
mN_eN_h}=\delta_{n\ell}\kappa_{nmN_eN_h}^2+V_{n\ell m},\\
&&V_{n\ell m}=\langle R_{nm}\vert\left(-\frac{2}{\rho}\right)\vert
R_{\ell
m}\rangle,\nonumber\\
&&d_{n m}=\langle R_{n m}\frac{e^{im\phi}}{\sqrt{2\pi}}\vert
M(\rho,\phi)\rangle\nonumber\\
&&=(M_0\rho_0)\frac{2\gamma}{\sqrt{\pi}}\sqrt{n+1}\frac{\left(1-\frac{\gamma\rho_0^2}{2}\right)^{n}}
{\left(1+\frac{\gamma\rho_0^2}{2}\right)^{n+2}}.\nonumber
\end{eqnarray}
Here we have used the dipole density $M(\rho,\phi,z_e,z_h)$ in the
form
\begin{equation}
M(\rho,\phi,z_e,z_h)=M_0\sqrt{\frac{2}{\pi}}\frac{\rho}{\rho_0^3}e^{-\rho^2/2\rho_0^2}\frac{e^{i\phi}+e^{-i\phi}}{\sqrt{2\pi}}
\delta(z_e-z_h),\end{equation} and
 \begin{eqnarray}\label{defknmFaraday}
&&\kappa_{n
mN_eN_h}^2=\frac{2\mu}{\hbar^2}a^{*2}(E_g-\hbar\omega-i{\mit\Gamma})\nonumber\\
&&+U_{n m}/R^*+\frac{W_{eN_e}+W_{hN_h}}{R^*},\\
 &&U_{n m}/R^*=\gamma\left(2n
+\hbox{sgn}\,(B)\;m\frac{\mu}{\mu'}+\vert m\vert
+1\right).\nonumber
\end{eqnarray}
The detailed form of the matrix elements
$V_{n\ell m}$ is given by Eq. (\ref{matrix_elem_HF}) in Appendix \ref{Appendix
A}. With the help of the coefficients
$c$ one can get the exciton amplitude $Y$, which is then substitutes into Eq.
(\ref{def_mean_susceptibility}), from which the mean
magneto-susceptibility for the case of high magnetic fields can be
determined. Restricting the considerations to the lowest
confinement state in the $z$-direction and denoting $\kappa_{n
m00}=\kappa_{nm}$, the magneto-susceptibility for the Faraday
configuration for the high field is given by the follwing formula
\begin{eqnarray}\label{chi_Faraday_HF}
&&\chi^F=\frac{16}{3\pi}\epsilon_b\gamma^2\left(\frac{a^*}{L}\right)\frac{\Delta_{LT}}{R^*}e^{4\rho_0}
\,\frac{\alpha_e\alpha_h}{p}\nonumber\\
&&\times\hbox{erf}\,\left(\frac{L\sqrt{p}}{2}\right)\sum\limits_{n=0}^N\sum_m\,c_{nm}d_{n1},\nonumber\\
&&d_{n1}=\sqrt{n+1}\exp[2\rho_0-(n+1)\gamma\rho_0^2],\\
&&\sum\limits_{n=0}^{n_{max}}a_{n\ell\,m}c_{nm}=d_{\ell\,1},\nonumber\\
&&a_{n\ell
m}=\delta_{n\ell}\kappa_{nm}^2+V_{n\ell},\nonumber\\
&&\kappa_{n m}^2=\frac{E_g-\hbar\omega-i{\mit\Gamma}+U_{n
m}+W_{e0}+W_{h0}}{R^*}.\nonumber
\end{eqnarray}
\subsection{Intermediate
fields}\label{Farady_intermediate} For intermediate magnetic
fields the exciton energies and the Landau states energies are
comparable, therefore we must include  the contributions from  Coulomb interaction
and the magnetic field at the same footing. 
In Ref.
\cite{Magnetoexcitons_2019} we have developed the method for such calculations and here we will recall its fundamental points. The
Eq. (\ref{Ylinear}) has to be transformed into a Lippmann-Schwinger equation
\begin{equation}\label{Lippmann1}
H_{kin+B+confinement}Y=ME-VY,
\end{equation}
where $V$ is the 2-dimensional Coulomb e-h interaction potential,
and $H_{kin+B+confinement}$ is the
"kinetic+magnetic+confinement\"\,part of the Hamiltonian
(\ref{SL3Hamiltonian}). The above equation can be solved by means
of an appropriate Green's function \cite{Mott}
\begin{equation}\label{Lippmann2}
Y=GME-GVY.\end{equation} The Green function has the form \cite{Mott}
\begin{eqnarray*}
&&G(\rho,\rho';\phi,\phi';z_e,z_e';z_h,z_h')=\nonumber\\
&&=\frac{1}{2\pi}\sum\limits_{N_e.N_h}\sum\limits_{n=0}^\infty\sum_m
e^{im(\phi-\phi')}\psi^{(1D)}_{\alpha_h,N_h}(z_h)\psi^{(1D)}_{\alpha_h,N_h}(z_h')\\
&&\times
\psi^{(1D)}_{\alpha_e,N_e}(z_e)\psi^{(1D)}_{\alpha_e,N_eh}(z_e')\frac{R_{nm}(\rho)R_{nm}(\rho')}
{\kappa_{nmN_eN_h}^2},\nonumber
\end{eqnarray*}
where $R_{nm}(\rho)$ are given in Eq. (\ref{magnetic_eigen_f}),
and $\kappa_{nmN_eN_h}^2$ is given by Eq. (\ref{defknmFaraday}).

The Lippmann-Schwinger equation (\ref{Lippmann1}) is an integral
equation for the unknown function $Y$. There are several methods
to solve such equations. We choose the method of an trial function
$Y$, which we take in the form
\begin{eqnarray}\label{trial_Faraday}
&&Y=\Psi_{00}R_{01}(\rho)\left[\sum\limits_{m=\pm
1}Y_{0m,00}\exp(-\kappa_{0m00}\rho)\frac{e^{im\phi}}{\sqrt{2\pi}}\right]\nonumber\\
&&+\sum\limits_{n=1}^\infty\sum\limits_{N_e,N_h\geq
1}\sum\limits_m \frac{e^{im\phi}}{\sqrt{2\pi}}Y_{nmN_eN_h}
R_{nm}(\rho)\Psi_{N_eN_h},
\end{eqnarray}
where $Y_{nmN_eN_h}$ are coefficients to be determined, and
\begin{equation}
\Psi_{N_eN_h}=\psi^{(1D)}_{\alpha_e,N_e}(z_e)\psi^{(1D)}_{\alpha_h,N_h}(z_h).
\end{equation}
The exciton amplitude $Y$, and thus the magneto-susceptibility, is
known once the parameters $Y_{nmN_eN_h}$ are calculated. The
method of calculation is given in Appendix \ref{Appendix B}, where
we obtained
\begin{eqnarray}\label{Y0Faraday}
&&Y_{0\pm 1,00}=\frac{2\mu}{\hbar^2
a^*}\left[(M_0\rho_0)\frac{2\gamma}{\sqrt{\pi}}\right]{\mathcal
E}\nonumber\\
&&\times\frac{d_{01}e^{-z^2/4}}{3\kappa_{0\pm
1,00}^2D_{-4}(z)-2\sqrt{\gamma}D_{-3}(z)},
\end{eqnarray}
where $D_\nu(z)$ are parabolic cylinder functions,\cite{Grad}
\begin{displaymath}
z=\frac{\kappa_{0\pm 1,00}}{\sqrt{\gamma}},
\end{displaymath}
and
\begin{eqnarray}\label{YnFaraday}
&&Y_{n\pm 1,00}=\frac{2\mu}{\hbar^2
a^*}\left[(M_0\rho_0)\frac{2\gamma}{\sqrt{\pi}}\right]{\mathcal
E}\frac{d_{n1}}{\kappa^2_{0\pm 1,00}}.\end{eqnarray} With the
above quantities, substituted in Eq. (\ref{trial_Faraday}) and Eq.
(\ref{def_mean_susceptibility}), we obtained the mean QW
magneto-susceptibility in the Faraday configuration and
intermediate magnetic field regime in the form
\begin{eqnarray}\label{chiF_med}
&&\chi=\frac{16}{3\pi}\epsilon_b\gamma^2\left(\frac{a^*}{L}\right)\frac{\Delta_{LT}}{R^*}
e^{4\rho_0}\frac{\alpha_e\alpha_h}{p}\hbox{erf}\,\left(\frac{L\sqrt{p}}{2}\right)\nonumber\\
&&\times \sum_{m=\pm
1}\Biggl\{\frac{3d_{0m}\exp\left(\frac{u^2}{4}\right)D_{-4}(u)}
{\exp(z^2/4)\left[3\kappa_{0m}^2D_{-4}(z)-2\sqrt{\gamma}D_{-3}(z)\right]}\nonumber\\
&&+
\sum\limits_{n=1}^N\frac{d_{nm}^2}{\kappa_{nm}^2}\Biggr\},\\
&& d_{nm}=d_{n\vert
m\vert}=\sqrt{n+1}\frac{(1-\gamma\rho_0^2/2)^{n}}{(1+\gamma\rho_0^2/2)^{n+2}}\nonumber\\&&\approx\sqrt{n+1}\;e^{-(n+1)\gamma\rho_0^2},\nonumber\\
&&u=\frac{\kappa_{0\pm 1}}{s},\quad
s=\frac{1}{\rho_0}\left(1+\frac{\gamma\rho_0^2}{2}\right)^{1/2},\;z=\frac{\kappa_{0\pm
1}}{\sqrt{\gamma}}.\nonumber
\end{eqnarray}

\section{The Voigt configuration}
In the Voigt configuration the magnetic
field is perpendicular to the wave vector of the propagating
electromagnetic wave and, in the QW geometry, parallel to the QW
planes.
\subsection{Weak field
regime}\label{Voigt_weak}  We choose the magnetic field  \textbf{B} parallel to the
$0x$-axis, which corresponds to the vector potential
\begin{equation}
\textbf{A}=\frac{B}{2}(0,-z,y).\end{equation} With this potential
and the confinement potentials (\ref{confinement}), the QW
Hamiltonian (\ref{BFmagnetichamiltonian1}) takes the form
\begin{eqnarray}\label{VoigtHamilt2}
&&H^V_{\rm QW}=E_g+\frac{1}{2m_e}p_{ex}^2+\frac{1}{2m_h}p_{hx}^2\nonumber\\
&&+\frac{1}{2m_e}p_{ey}^2+\frac{1}{2m_h}p_{hy}^2+\frac{1}{8m_e}e^2B^2y_e^2+\frac{1}{8m_h}e^2B^2y_h^2\nonumber\\
&&-\frac{1}{2m_e}p_{ey}eBz_e
+ \frac{1}{2m_h}p_{hy}eBz_h\\
&&+\frac{1}{2m_e}p_{ez}^2+\frac{1}{8m_e}e^2B^2z_e^2 +\frac{1}{2}m_{e}\omega_{ez}^2z_e^2\nonumber\\
&&+\frac{1}{2m_h}p_{hz}^2
 +\frac{1}{2}m_{hz}\omega_{hz}^2z_h^2+\frac{1}{8m_h}e^2B^2z_h^2\nonumber\\
&& +p_{ez}eBy_e-p_{hz}eBy_h\nonumber\\
&&
-\frac{e^2}{4\pi\epsilon_0\epsilon_b}
\left[\left(x_e-x_h\right)^2+\left(y_e-y_h\right)^2\right]^{-1/2}.\nonumber
\end{eqnarray}
As in the case of the Faraday configuration, we will discuss the
three regimes: weak, intermediate and the high magnetic field with the proper form of the Hamiltonian for each of them.

In
the weak field limit we transform the Hamiltonian
(\ref{VoigtHamilt2}) to the form

\begin{eqnarray}\label{HQWV1}
&&H^V_{\rm
QW}=E_g+\frac{1}{2m_e}p_{ex}^2+\frac{1}{2m_e}p_{ey}^2\nonumber\\
&&-
\frac{e^2}{4\pi\epsilon_0\epsilon_b}\left[\left(x_e-x_h\right)^2+\left(y_e-y_h\right)^2\right]^{-1/2}\nonumber\\
&&+\frac{1}{2m_e}p_{ez}^2+\frac{1}{2}m_e\Omega_{ez}^2+\frac{1}{2m_h}p_{hz}^2+\frac{1}{2}m_h\Omega_{hz}^2+H',\nonumber\\
&&\Omega_{ez}^2=
\frac{\omega_{ec}^2}{4}+\omega_{ez}^2,\quad\omega_{ec}=\frac{eB}{m_e},\\
&&\Omega_{hz}^2=
\frac{\omega_{hc}^2}{4}+\omega_{hz}^2,\quad\omega_{hc}=\frac{eB}{m_h},\nonumber\\
&&H'=\frac{1}{8m_e}e^2B^2y_e^2+\frac{1}{8m_h}e^2B^2y_h^2.\nonumber
\end{eqnarray}
Introducing the relative and the center-of-mass coordinates $y,M_Y$
in the $y$ direction
\begin{displaymath}
M_Y=\frac{m_ey_e+m_hy_h}{M},\quad y=y_e-y_h,\qquad M=m_e+m_h
\end{displaymath}
we transform the Hamiltonian (\ref{HQWV1}) to the form
\begin{eqnarray}\label{Hamiltonian_Voigt}
&&H^V_{\rm
QW}=E_g+H^{(2D)}_{Coul}+H^{(1D)}_{m_{e},\Omega_{ez}}\left(z_e\right)+H^{(1D)}_{m_{h},\Omega_{hz}}\left(z_h\right)+H',\nonumber\\
&&H'=\frac{1}{8\mu}e^2B^2M_Y^2+\frac{1}{8\mu}e^2B^2y^2q+\frac{1}{4\mu'}e^2B^2M_Yy,\\
&&q=\frac{m_h^2-m_h m_e+m_e^2}{M^2}.\nonumber
\end{eqnarray}
We will proceed in a similar way as in the case of a weak field in
the Faraday configuration. Treating the magnetic part  $H'$ as a
perturbation, we assume the solution for $Y$ in the form
(\ref{expansion}), with the eigenfunctions appropriate to the
Hamiltonian (\ref{Hamiltonian_Voigt}).  This leads to the system
of equations (\ref{system_weak_F}) for the expansion coefficients
$c$, where now the matrix elements are given by
\begin{equation}\label{matrix_elem_Voigt}
V^{Voigt}_{jkm}=\frac{1}{2}q\,V^{Faraday}_{jkm},\end{equation}
with $V^{Faraday}_{jkm}$ defined in Eq. (\ref{M_Elem_weak}).

 The susceptibility is
obtained in the form (\ref{chiF_weak}), where
\begin{eqnarray}\label{PsiNNVoigt}
&&\langle
\Psi_{NN}\rangle_L=\frac{1}{2^N\,N!}\frac{2}{\sqrt{\pi}}\int\limits_0^{\alpha^V
L/2}e^{-t^2}H_N^2(t)dt,\\
&&
\alpha_e^V=\alpha_h^V=:\alpha^V=\frac{1}{a^*}\left[\frac{\gamma^2}{4}+\left(\frac{\pi
a^*}{L}\right)^4\right]^{1/4},\nonumber\\
&&\frac{W_{e0}^V}{R^*}+\frac{W_{h0}^V}{R^*}=\frac{W^V_{eh0}}{R^*}=\left[\frac{\gamma^2}{4}+\left(\frac{\pi
a^*}{L}\right)^4\right]^{1/2},\nonumber\end{eqnarray} and, for
$N_e=N_h=N$, the confinement energies are now defined as

\begin{eqnarray}\label{confinement_Voigt}
&&W^V_{eN}+W^V_{hN}=W_{NN}=(2N+1)W^V_{eh0},\nonumber\\
&& N=0,1,2,\ldots,N_{max},\\
&&W^V_{eh0}=\left[\frac{\gamma^2}{4}+\left(\frac{\pi
a^*}{L}\right)^4\right]^{1/2} R^*.\nonumber
\end{eqnarray}
\subsection{ High field regime}\label{Voigt_high}
In the high field limit for the Voigt configuration the e-h
Coulomb interaction is considered as a perturbation, so the
unperturbed QW Hamiltonian has the form
\begin{eqnarray}\label{HQWV}
&&H^V_{QW}=\frac{p_x^2}{2\mu}+H^{(1D)}_{\mu,\Omega_y}(y)+H^{(1D)}_{m_e,\Omega{ez}}(z_e)+H_{m_h,\Omega_{hz}}(z_h),\nonumber\\
&&\frac{\hbar\Omega_y}{2R^*}=\frac{\gamma}{2}\sqrt{q},
\end{eqnarray}
with $q$ defined in Eq.(\ref{Hamiltonian_Voigt}). We apply the
method of the so called adiabatic potentials, used in bulk
crystals (see\cite{Magnetoexcitons_2019} and references therein),
here adapted for the case of QWs. The exciton amplitude $Y$ will
be assumed in the form
\begin{eqnarray}\label{expansion_V_HF}
&&Y(x,y,z_e,z_h)\\
&&=\sum\limits_{N_xN_yN_eN_h}c_{N_xN_yN_eN_h}\psi_{N_x}(x)\psi^{(1D)}_{\beta,\Omega_y}(y)\Psi_{N_eN_h}(z_e,z_h),\nonumber
\end{eqnarray}
where
\begin{eqnarray}\label{psi1Dbeta}
&&\psi^{(1D)}_{\beta,N_y}=\pi^{-1/4}\sqrt{\frac{\beta}{2^{N_y}N_y!}}H_{N_y}(\beta
y)e^{-\beta^2y^2/2},\nonumber\\
&&\beta=\frac{1}{a^*}\sqrt{\frac{\hbar\Omega_y}{2R^*}}
=\frac{1}{a^*}\,q^{1/4}\sqrt{\frac{\gamma}{2}}=\frac{1}{a^*}\tilde{\beta},\\
&&\Psi_{N_eN_h}(z_e,z_h)=\psi^{(1D)}_{\alpha_e^V,N_e}(z_e)\psi^{(1D)}_{\alpha_h^V,N_e}(z_h),\nonumber
\end{eqnarray}
and $\psi_{N_x}(x)$ are eigenfunctions of the operator
\begin{equation}\label{adiab1}
H_x=\frac{p_x^2}{2\mu}+V_{N_yN_y'}(x),\end{equation} where
\begin{equation}\label{adiab2}
V_{N_yN_y'}(x)=-2\int\limits_{-\infty}^\infty dy
\frac{\psi^{(1D)}_{\beta,N_y}(y)\psi^{(1D)}_{\beta,N'_y}(y)}{\sqrt{x^2+y^2}}.
\end{equation}
We restrict the discussion to the diagonal terms $V_{N_yN_y}$, and
approximate the expression (\ref{adiab2}) by
\begin{equation}
V_{N_y}=-\frac{2}{a_{N_y}+\vert x\vert}.
\end{equation}
The coefficients $a_{N_y}$, for odd parity eigenfunctions
$\psi^{(1D)}_{\beta,N_y}, \;N_y=2n+1$, are calculated in Appendix \ref{Appendix
C}. In this approximation the Schr\"{o}dinger equation with the
operator (\ref{adiab1}) becomes
\begin{equation}
\left(\frac{p_x^2}{2\mu}-\frac{2}{a_{N_y}+\vert
x\vert}\right)\psi=E\psi,
\end{equation}
which gives the eigenfunctions
\begin{equation}
\psi_{jn}(x)=\frac{\sqrt{2}}{j+1}e^{-(\vert x\vert
+a_{2n+1})/(j+1)}L^1_{j}\left[\frac{2(\vert x\vert
+a_{2n+1})}{j+1}\right],
\end{equation}
$j=0,1,\ldots,$ and eigenvalues
\begin{equation}
E_j=-\frac{R^*}{(j+1)^2}.\end{equation} Having the above
functions, and using the dipole density in the form
\begin{eqnarray}
&&M(x,y,z_e,z_h)\\
&&=\frac{M_0}{2\rho_0^3}\sqrt{\frac{2}{\pi}}
e^{-\frac{x^2}{2\rho_0^2}}y\,e^{-\frac{y^2}{2\rho_0^2}}\delta(z_e-z_h),\nonumber\end{eqnarray}
 we calculate
the expansion coefficients in the formula Eq.
(\ref{expansion_V_HF}) and thus the exciton amplitude $Y$ and,
finally, the mean QW magneto-susceptibility for the Voigt
configuration in the limit of high magnetic fields

\begin{eqnarray}\label{chiv_high}
&&\chi^V=
\frac{4\sqrt{\pi}}{3}\epsilon_b\Delta_{LT}e^{4\rho_0}\sum\limits_{j=0}^{N_xmax}
\sum\limits_{n=0}^{N_ymax}\sum\limits_{N=0}^{N_zmax}
\frac{2}{(j+1)^2}\,\nonumber\\
&&\times\,e^{-\frac{2a_{2n+1}}{j+1}}\left[L_{j}^{(1)}\left(\frac{2\,a_{2n+1}}{j+1}\right)\right]^2\\
&&\times\left(\frac{2\tilde{\beta}}{1+\tilde{\beta}^2\rho_0^2}\right)^{3}\frac{(2n+1)!}{2^{2n+1}(n!)^2}
\left(\frac{1-\tilde{\beta}^2\rho_0^2}{1+\tilde{\beta}^2\rho_0^2}\right)^{2n}\nonumber\\
&&\times\left[E_g-\hbar\omega-\frac{R^*}{(j+1)^2}+\left(2n+\frac{3}{2}\right)
\hbar\Omega_y+W_{NN}\right]^{-1},\nonumber
\end{eqnarray}
 where $\langle
\Psi_{NN}\rangle_L$ are defined in Eq. (\ref{PsiNNVoigt}), and the
confinement energies $W_{NN}$ in Eq. (\ref{confinement_Voigt}).
\subsection{ Intermediate
fields}\label{Voigt_intermediate} We calculate the mean
magneto-susceptibility for the Voigt configuration and in the
regime of intermediate magnetic fields by the Green function
method described above for the case of Faraday configuration.
Again, we use the Lippmann-Schwinger equation (\ref{Lippmann2}) to
calculate the exciton amplitude $Y$, which is then used to obtain
the magneto-susceptibility. {The Green's function in Eq.
(\ref{Lippmann2}) satisfies, by definition, the equation
\begin{eqnarray*}
&&H^VG(x,x';y,y';z_e,z_e';z_h,z_h')\\
&&=-\delta(x-x')\delta(y-y')\delta(z_e-z_e')\delta(z_h-z_h')
\end{eqnarray*}
where the operator $H^V$ has the form (\ref{HQWV}). Expressing
Green's function in terms of eigenfunctions of the operators
contained in $H^V$ one obtains
\begin{eqnarray}\label{Green_V_intermediate}
&&G=\frac{2\mu}{\hbar^2}\sum\limits_{n,N_e,N_h}{\frac{1}{2\pi}}\int\limits_{-\infty}^\infty
dk\,e^{ik(x-x')}\psi^{(1D)}_{\beta,n}(y)\psi^{(1D)}_{\beta,n}(y')\nonumber\\
&&\times
\frac{\psi^{(1D)}_{\alpha_e^V,N_e}(z_e)\psi^{(1D)}_{\alpha_e^V,N_e}(z_e')
\psi^{(1D)}_{\alpha_h^V,N_h}(z_h)\psi^{(1D)}_{\alpha_h^V,N_h}(z'_h)}{k^2+\kappa_{nN_eN_h}^2}
\end{eqnarray}
 with
\begin{eqnarray}\label{defknm2}
&&\kappa_{nN_eN_h
}^2=\frac{2\mu}{\hbar^2}\Biggl[(E_g-\hbar\omega-i{\mit\Gamma})\\
&&+\left(2n+\frac{3}{2}\right)\hbar\Omega_y+
\left(N_e+\frac{1}{2}\right)\hbar\Omega_{ez}+\left(N_h+\frac{1}{2}\right)\hbar\Omega_{hz}\Biggr].\nonumber
\end{eqnarray}
The functions $\psi^{(1D)}_{\beta,n}(y)$ are defined in Eq.
(\ref{psi1Dbeta}). For the further calculations we must specify a
trial function $Y$. Accounting only the lowest confinement state
we use the following trial function of the form
\begin{eqnarray}\label{expansionV}
&&Y=Y_{0}\Psi_{00}\psi_{1,\beta}^{(1D)}(y)
e^{-\kappa_{0}\sqrt{x^2+y^2}} \\&&+\sum\limits_{n=1}^\infty
\sum\limits_{N_eN_h\geq
1}\psi_{2n+1,\beta}^{(1D)}(y)\Psi_{N_eN_h}\frac{1}{2\pi}\int\limits_{-\infty}^\infty
dk\,Y_{nN_eN_h}(k) e^{ikx},\nonumber
\end{eqnarray}
 where
$\kappa_0^2=\kappa_{000}^2$, $Y_0, Y_{nN_eN_h}$ $\Psi_{N_eN_h}$ is defined in Eq.
(\ref{psi1Dbeta}) and  coefficients have to be determined; the detailed calculations are presented  in
Appendix D. With the help of these the coefficients  determined the  the exciton
amplitude and than, similary as in  the section III C one can calculate the mean magneto-susceptibility for the
Voigt configuration and in the intermediate field regime, arriving to the formula
\begin{eqnarray}\label{chiVintermediate}
&&\chi^{intermV}=\nonumber\\
&&=\frac{4}{3}\frac{\Delta_{LT}}{R^*}\epsilon_b\left(\frac{a^*}{L}\right)\langle
\Psi_{00}\rangle_L\Biggl\{
\frac{1}{\sqrt{\pi}\kappa_0}\rho_0^3\left(\frac{2\beta}{1+\beta^2\rho_0^2}\right)^
{3}\nonumber\\
&&\times\exp\left[\frac{\kappa_0^2\rho_0^2}{4(1+\beta^2\rho_0^2)}\right]
D_{-3}\left(\frac{\kappa_0\rho_0}{\sqrt{1+\beta^2\rho_0^2}}\right)\nonumber\\
&&\times\left[\frac{2^{3/2}}{\sqrt{\pi}}e^{\kappa_0^2/8\beta^2}D_{-3}\left(\frac{\kappa_0}{\beta\sqrt{2}}\right)
-F(\kappa_0,\beta)\right]^{-1}\nonumber\\
&&+\sum\limits_{n\geq 1,N\geq 1}\Biggl\{
\langle\Psi_{NN}\rangle_L\frac{1}{2^{2n}}\\
&&\times\left(\frac{\tilde{\beta}}{1+\tilde{\beta}^2\rho_0^2}\right)^3
\frac{(2n+1)!}{(n!)^2}\left(\frac{\tilde{\beta}^2\rho_0^2-1}{\tilde{\beta}^2\rho_0^2+1}\right)^{2n}\frac{\pi}
{\kappa_{nN}}w(i\beta\,\kappa_{nN})\Biggr\}\nonumber,
\end{eqnarray}
$W_{\kappa,\mu}(z)$ is Whittaker's function of the second kind,
$w(z)$ is the complex error function,\cite{Abramowitz} and
$F(\kappa_0,\beta)$ is defined in Eq. (\ref{Fkappa}).

\begin{table}[ht!]
\caption{\small Band parameter values for Cu$_2$O, masses in free
electron mass $m_0$, $R^*$ calculated from
$(\mu/\epsilon_b^2)\cdot 13600\,\hbox{meV}$,\,
$R^*_{e,h}=(m_{e,h}/\mu)R^*, a^*_{e,h}=(\mu/m_{e,h})a^*$}
\begin{center}
\begin{tabular}{p{.2\linewidth} p{.2\linewidth} p{.2\linewidth} p{.2\linewidth} p{.2\linewidth}}
\hline
Parameter & Value &Unit&Reference\\
\hline $E_g$ & 2172.08& meV& \cite{Kazimierczuk}\\
$R^*$&87.78& meV &\\
$\Delta_{LT}$&$1.25\times 10^{-3}$&{meV}& \cite{Stolz}\\
$m_e$ & 0.99& $m_0$&\cite{Naka}\\
$m_h$ &0.58&  $m_0$&\cite{Naka}\\
$\mu$ & 0.363 &$m_0$&\\
$M_{tot}$&1.56& $m_0$&\\
$a^*$&1.1& nm&\cite{Kazimierczuk}\\
$r_0$&0.22& nm&\cite{Zielinska.PRB}\\
$\epsilon_b$&7.5 &&\cite{Kazimierczuk}\\
${R_e^*}$&239.4&meV&\\
${R^*_h}$&140.25&meV&\\
 ${a^*_e}$&0.4 &nm&\\
 ${a^*_h}$&0.69 &nm&\\
 $\Gamma_j$&3.88/$j^3$ &meV&\cite{Kazimierczuk,maser2}\\
 \hline
\end{tabular} \label{parametervalues}\end{center}
\end{table}

\section{Results of specific calculations}\label{Results}
We have calculated the QW magneto-absorption from
the imaginary part of the magneto-susceptibilities, given for the
Faraday configuration in equations (\ref{chiF_weak}),
(\ref{chi_Faraday_HF}), (\ref{chiF_med}), and for the Voigt
configuration in equations (\ref{chiF_weak}) (with adequate change
of parameters), (\ref{chiv_high}), and (\ref{chiVintermediate}).
The parameters used in calculations are collected in  Table
\ref{parametervalues}. We assume that the QW band parameters (for
example, effective masses), are equal to their bulk values.
 Since the quantum well thickness under consideration is $L\ge 20$ nm, it is much larger than the exciton (1.1 nm for n=1, see Ref\cite{Kazimierczuk}), the choice of bulk effective masses is  justified. 
The calculations have been performed for the whole magnetic field
strength spectrum, including the weak, intermediate, and high
field regimes.

\subsection{Estimation of regime boundaries}
The problem of delimiting boundaries of magnetic fields regimes   requires specific analysis for each material. Below we will present a heuristic and  simple  method, which allows for rough estimation of these limits.
The lowest Landau energies for
p-exciton (including the Zeeman splitting) given by (see (Eq.
\ref{defknmFaraday}))
\begin{equation}
U_{0,\pm 1}=\left(\frac{B}{B_{cr}}\right)\left(2\pm
\frac{\mu}{\mu'}\right)R^*,\end{equation} are compared to the
2-dimensional hydrogen energy, which for $n=1,\;m=\pm 1$ is equal to $4R^*/25$, thus the equation 
%\begin{equation}\label{limit1}
%\frac{B}{B_{cr}}=\frac{4}{9\left(2\pm \frac{\mu}{\mu'}\right)}
%=\gamma_{cr}
%\end{equation}
\begin{equation}\label{limit2}
\frac{B}{B_{cr}}=\frac{4}{25\left(2\pm \frac{\mu}{\mu'}\right)}=\gamma_{cr}.
\end{equation}
The parameter $\gamma_{cr}$ determines the limit of the weak field: for $B<\gamma_{cr}B_{cr}$ one
deals with the weak field;  $B\geq \gamma_{cr}B_{cr}$ indicates
 the intermediate field regime. For the Cu$_2$O data from Table
\ref{parametervalues}, depending on the quantum number $m$, we
obtain the limiting values 26.8 T and 29.2 T. The upper value
corresponds to $m=-1$ and the lower one to $m=1$. The limiting values
of the field decrease with increasing the Landau state number $n$.

The limits of the high field regime in
the Faraday configuration can be estimated using the matrix
elements given in Eq. (\ref{coeffHF}). Recalling parameter
$\gamma$ given by Eq. (\ref{gamma}) and comparing the Landau
energy (see Eq. (\ref{limit2})) with the value
of the matrix elements
\begin{eqnarray*}
&&W_{11}\approx 4\gamma,\quad \vert V_{11}\vert
=1.097\sqrt{\gamma},
\end{eqnarray*}
we obtain the critical value $\gamma_{cr}=0.075$,
which corresponds to the field strength $B$ is about 60 T. Note that this
evaluation can be interpreted only as a rough estimation; the real
positions of resonances are obtained by solving systems of
equations.

 In the Voigt configuration, the limits of the weak and
intermediate fields, can be derived in the same as in the Faraday
configuration. For the weak field we use the expression (42) and
compare with the unperturbed energy values. The Voigt matrix
elements are smaller than these for the Faraday one, for two
reasons. First, in this configuration the magnetic field
influences  only on the one degree of freedom.\cite{RivistaGC} Additionally, the  factor $q$ (Eq. 41), depending on the
effective electron and hole masses $v=m_e/m_h$, plays an
important role. The function $q(v)$ attains values: 1 for $v\to 0,
v\to\infty$ and attains its minimal value for $v=1$ (as in  "the positronium
model"). For Cu$_2$O ($v=1.429$) the parameter $q =0.273$ approaches close to that of positronium, so
 taking the Landau state $n=1$ and the matrix element
$V_{11}$, one obtains the limiting value $\gamma_{cr}$
corresponding to the field strength $B=56\,\hbox{T}$. Comparing
this value with the above indicated limiting values for the
Faraday configuration we see, that the limiting values defining
the weak field regime are about two times larger for the Voigt configuration than in the Faraday case. Other related physical effect
is that the field-induced blue shift of resonances in the Voigt
configuration is much smaller than that in the Faraday
configuration; which was experimentally observed and this was  confirmed.\cite{Wang,Chernenko} It should be also pointed out  that when comparing
the Cu$_2$O magneto-optical spectra with spectra of other
semiconductors, that most of them have the $q$ value much larger
than Cu$_2$O (i.e., for  GaAs $q$ is almost 4 times larger).

 The high field limit for the Voigt
configuration will be obtained from comparison of the Landau
energies, which now have the form
\begin{eqnarray*}
&&E_n^V=\left(2n+\frac{3}{2}\right)\hbar\Omega_y,\\
&&\frac{\hbar\Omega_y}{2R^*}=\frac{\gamma}{2}\sqrt{q}=0.261
\gamma,\\
&&\hbar\Omega_y=0.523\,\gamma\,R^*,
\end{eqnarray*}
with the 2-dimensional excitonic energies. For the lowest exciton
energies $-(4/9)$ we obtain the critical magnetic field
strengths above 180 T.

As it was mentioned above, we are aware that presented method enables for only quantitative estimations  but, as it will be shown below, the use of parameter $\gamma$ evaluated in such a way, gives a good agreement with available experimental data. With all the above comments, we present the obtained results.

\subsection{Discussion of numerical calculations}
 The Fig.
\ref{fig1} depicts the absorption spectrum of a Cu$_2$O quantum
well in the Faraday configuration calculated for a range of field
strengths B=0-100 T and thickness $L=20$ nm.
\begin{figure}[ht!]
\includegraphics[width=.9\linewidth]{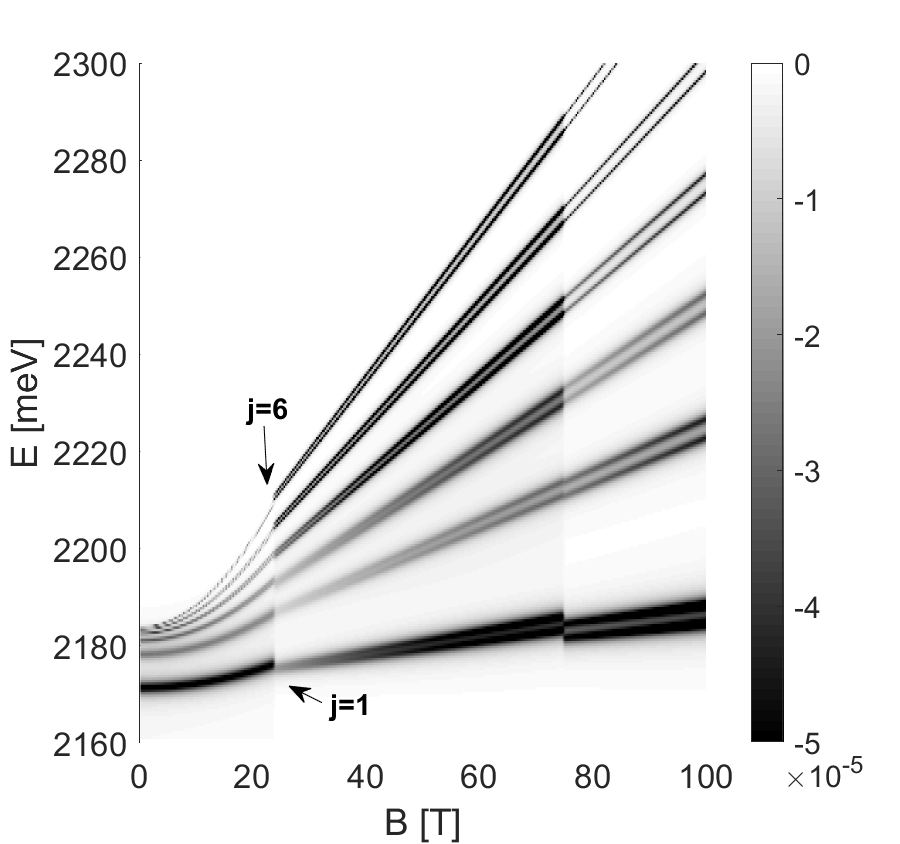}
\caption{Imaginary part of susceptibility (black color) of a quantum well in Faraday configuration, calculated in three field regimes. The lower limits of intermediate and high field regime are 28 T and 75 T accordingly}
\label{fig1}
\end{figure}
The boundaries between low, intermediate and high field regimes are estimated at
28 T and 75 T, which is a close match to the initial estimation of 26.8/29.2 and 74.49/81.36 T for $m\pm1$, $n=1$ respectively. For such values, there is a very good
correspondence between solutions (Eqs. (\ref{chiF_weak}),
(\ref{chiF_med}), (\ref{chi_Faraday_HF}) respectively). The lines appear in pairs,
corresponding to m=$\pm 1$ and exhibit roughly quadratic energy
shift with increasing $B$ in a weak field limit. Such tendency has been also observed in a bulk samples in the experiments and theoretically \cite{Rommel},\cite{Zielinska.PRB.2016.c}. Due to the fact that the energy shift of all lines is almost linear for B>10 T, the fit is not sensitive to the changes of low, medium and high field boundaries, so that even rough estimations presented above are sufficient to obtain continuous spectrum.

The Fig. \ref{fig2a} depicts the low field solution calculated for a bigger range of quantum numbers $[N,j,m]$, which are given in brackets. It is worth underscore that for QW with REs in a magnetic field these indexes describe three types of of states and three origins of resonances; $N$ arises from confinement in z-direction, number $j$ enumerating excitons is connected with e-h Coulomb interaction and $m$ refers to an interaction with magnetic field resulting with Zeeman splitting with
 lines shift towards higher energy with increasing field strength.One can observe several interesting tendencies. By increasing $N$, one introduces almost constant energy shift (series of blue lines for $j=2$, orange lines for $j=3$).
  On the other hand, the lines coming from higher excitonic states (red series) exhibit stronger energy shift with increasing $B$ due to bigger sensitivity of  higher  states to an external field and  finally, the split with respect to $m=\pm1$ is weaker for the higher $j$ lines. One has to do with an intricate situation of an interplay between Coulomb and magnetic interaction. 

\begin{figure*}[ht!]
\includegraphics[width=.7\linewidth]{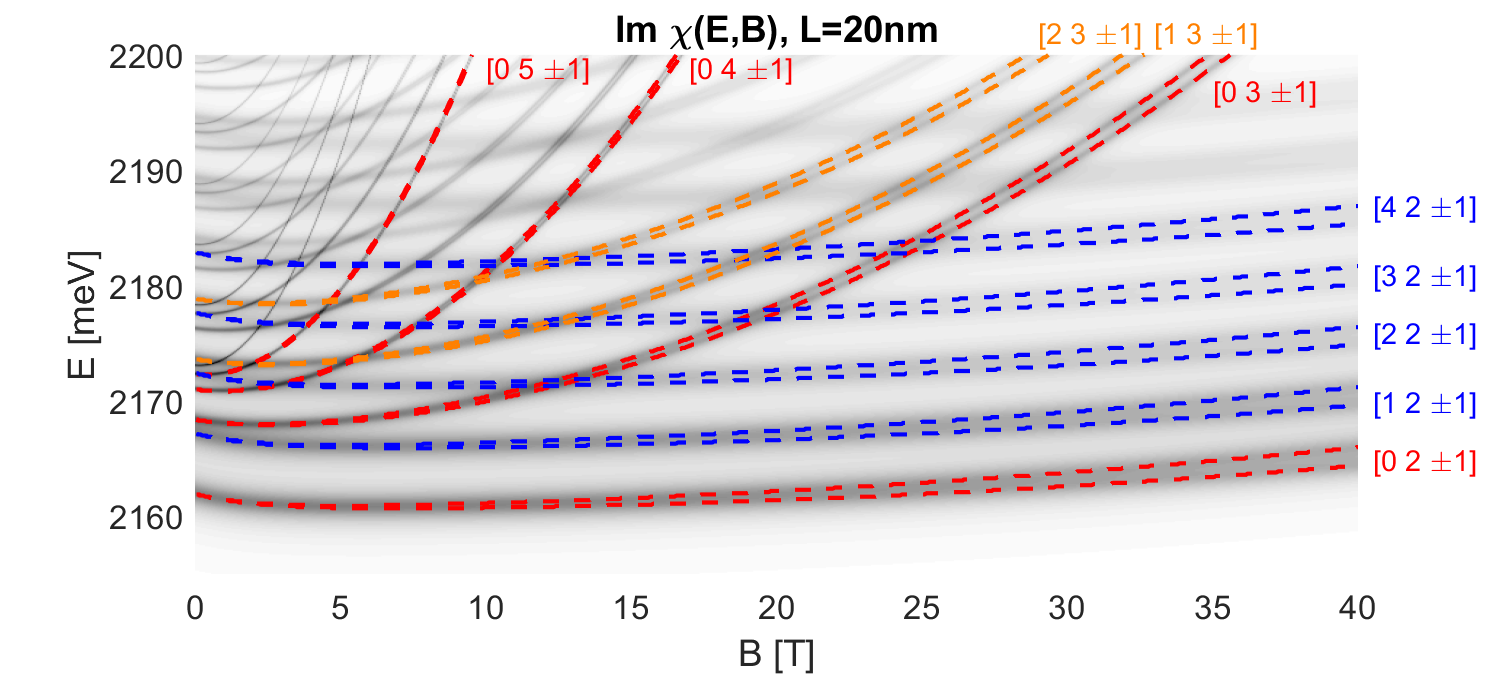}
\caption{The same as in Fig. \ref{fig1}, calculated for L=20 nm. The brackets denote quantum numbers $[N,j,m]$.}
\label{fig2a}
\end{figure*}

On the Fig. \ref{fig2b} one can observe the dependence of energy shift on the well thickness $L$. As one has expected the confinement effect is more pronounced for narrower QWs.  The states with various $N$ (blue lines) split from the respective $j$ state and diverge as $L\rightarrow 0$, with the higher $N$ states approaching $E\rightarrow\infty$ faster due to their lower binding energy and larger physical size, which makes them more affected by finite well size.  Higher $N$ states have are more affected by the potential barrier at the quantum well edges.  One can observe that the lines with different $j$ (red series) react to the confinement in the same manner - the distance between them remains almost constant up to $L \sim 5$ nm, where the well thickness becomes comparable to the exciton size. The large distance between $j=2$ and $j=3$ states is a result of high magnetic field ($B=50$ T); as mentioned before, lines with different $j$ exhibit different  energy shift depending on $B$, which results in increasing distance between them.

\begin{figure*}[ht!]
\includegraphics[width=.7\linewidth]{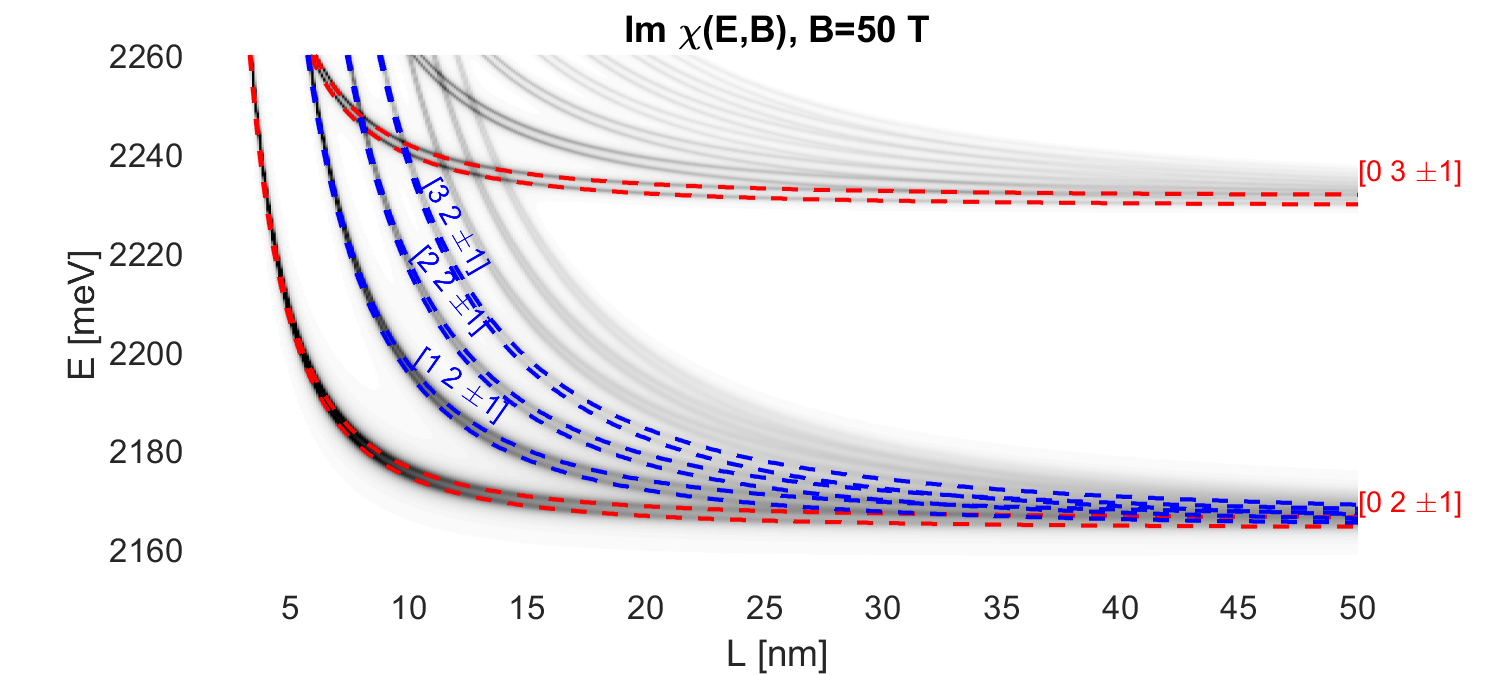}
\caption{The same as in Fig. \ref{fig1}, calculated for B=20 T. The brackets denote quantum numbers $[N,j,m]$.}
\label{fig2b}
\end{figure*}

The absorption spectrum in Voigt configuration appears to have a more complicated structure. The Fig. \ref{fig3} shows absorption coefficient calculated from Eq. (\ref{chiF_weak}) with Eq. (\ref{PsiNNVoigt}) for the weak regime Eq. (\ref{chiVintermediate}) for intermediate regime and from Eq. (\ref{chiv_high}) for the strong regime. The boundaries between regimes are set to 55 T and 140 T. The lower field limit is equal to the initial estimation and the high field limit is somewhat lower than initial estimation (180 T), but its exact location is very flexible due to the fact that energy shifts in both intermediate and strong field solutions are linear. Again, the fit between two regimes is the best for higher energy states. The most striking feature of the spectrum is the grouping of lines corresponding to the same value of $m$ which has the largest contribution to the state energy, especially in the high field regime. The energy shift depending on other quantum numbers ($N$ and $j$) is less pronounced, so that there are groups of lines centered around specific value of $m$. 

\begin{figure}[ht!]
\includegraphics[width=.9\linewidth]{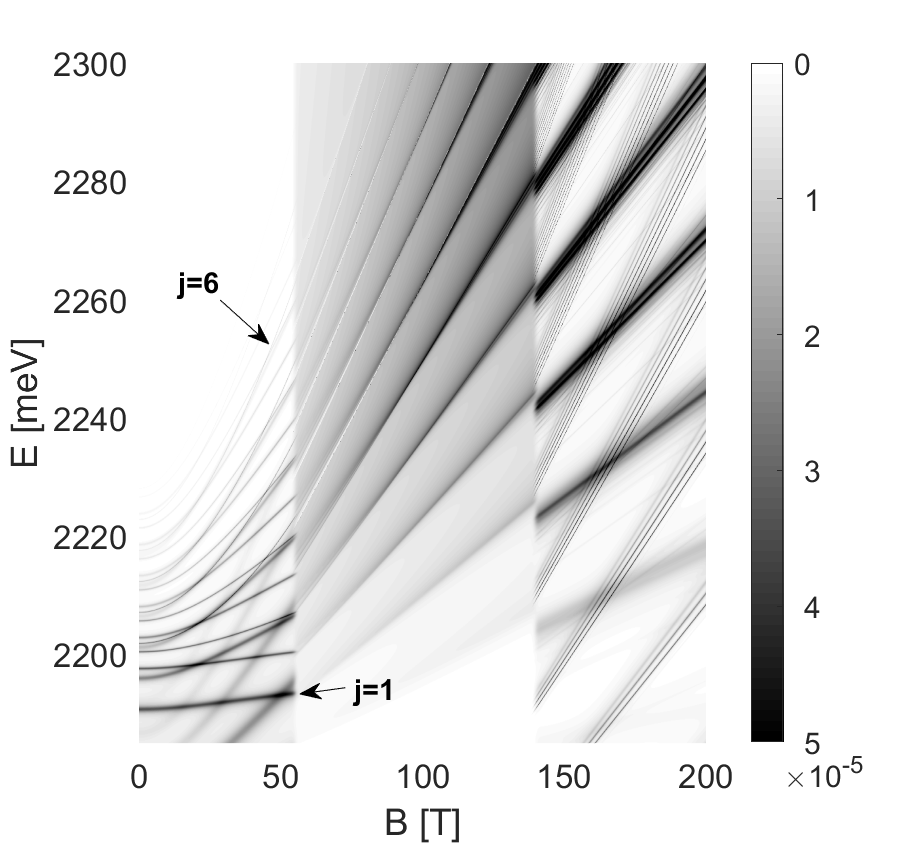}
\caption{Imaginary part of susceptibility of a quantum well in Voigt configuration calculated for a range of magnetic field strength and $L$=20 nm.}
\label{fig3}
\end{figure}

To better discern these states, one can assign the quantum numbers $[N,j,m]$ to them, as shown on the Fig. \ref{fig4}.
\begin{figure*}[ht!]
\includegraphics[width=.7\linewidth]{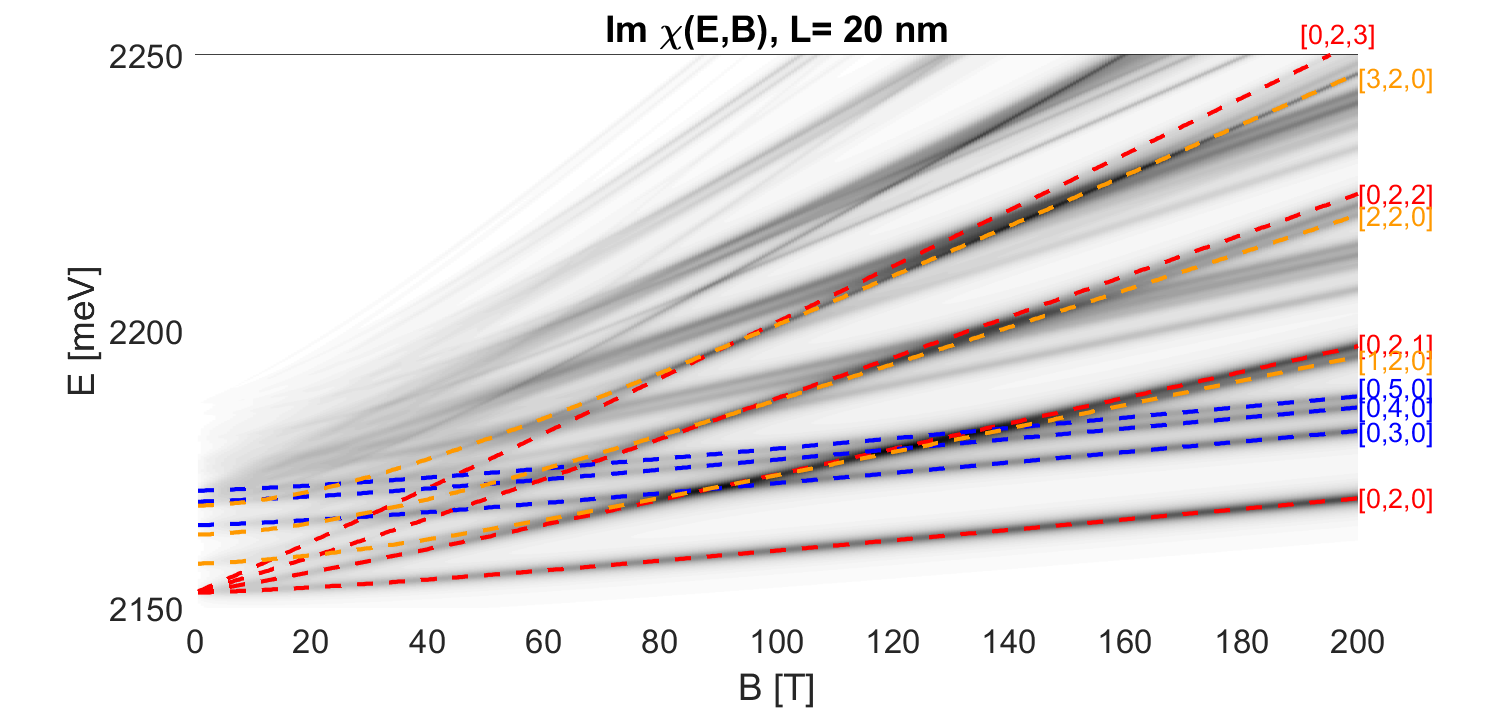}
\caption{The same as Fig. \ref{fig3}, with states identified by their quantum numbers [N,j,m].}
\label{fig4}
\end{figure*}
The base state, marked by red line, is $[0,2,0]$ and the other states are created by changing one quantum number. The increase of $j$ (blue lines) yields a typical, excitonic $\sim 1/n^2$ energy shift, approaching $E=E_g$ at $B=0$. In our model, the distance between excitonic states is independent of $B$. On the other hand, the energy shift with B depends strongly on $N$ and $m$. Every confinement state $N$ undergoes Zeeman split; one can see that the energy of $N=0$ states with various $m$ (red lines) changes linearly with B and these lines start from a common origin at $B=0$. The energy shift for higher $N$ (orange lines) is quadratic in the low field regime, and then transitions to linear at $B \sim 50$ T.

The dependence on the well thickness, shown on the Fig. \ref{fig5}, is also interesting. One can see that similar to the Faraday configuration, the energies diverge at the very low $L$ limit, with the exact location of the asymptote dependent on the quantum numbers $N$ and $j$ due to the fact that the physical size of exciton of any given $j$ affects the energy and the magnetic moment of a bound state depends on quantum number $N$.
\begin{figure*}[ht!]
\includegraphics[width=.7\linewidth]{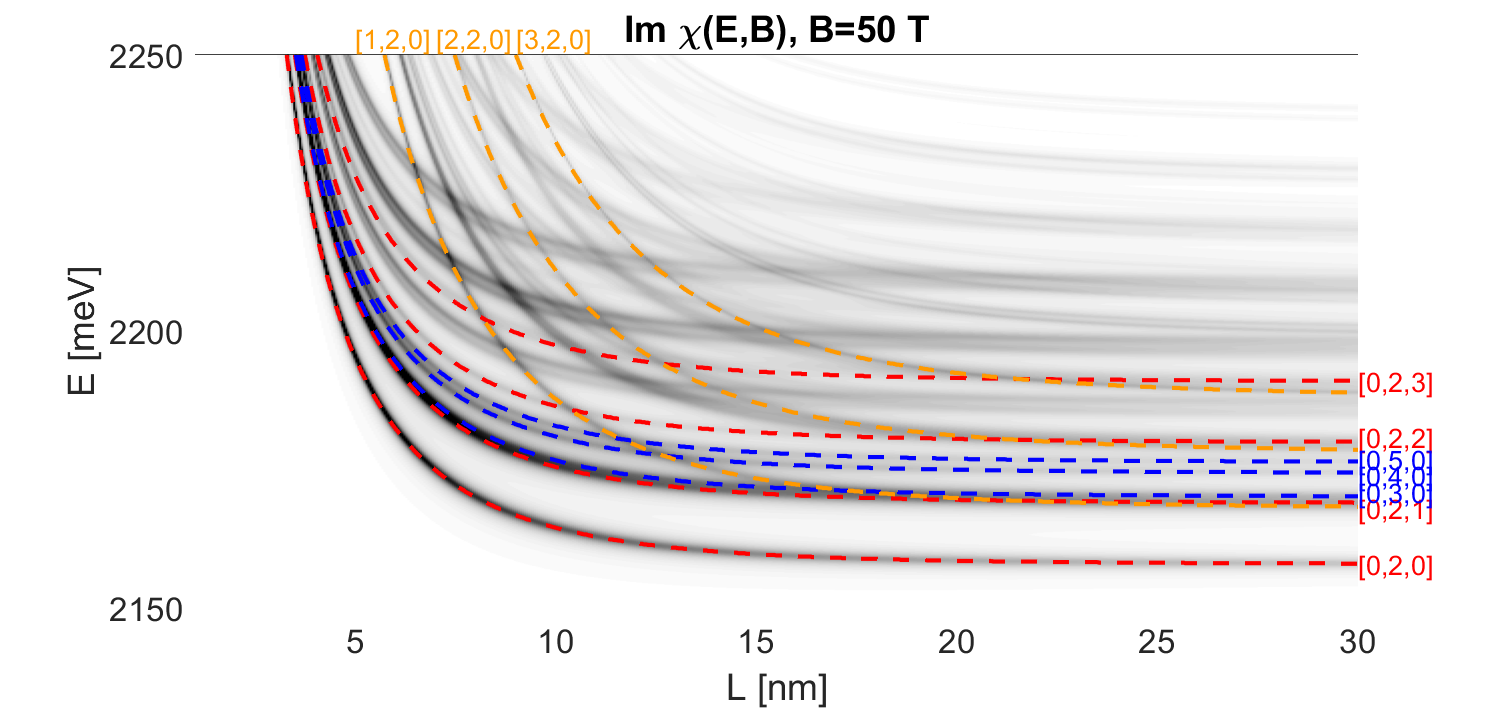}
\caption{The same as Fig. \ref{fig3}, calculated for $B=50 T$ and a range of $L$ values, with states identified by their quantum numbers [N,j,m].}
\label{fig5}
\end{figure*}

We also have performed the comparison of our theoretical results with available experimental data to verify the accuracy and applicability of our theoretical approach and estimations.

The Fig. \ref{fig6} shows a comparison between the energy of first confinement state measured in GaAlAs quantum dot in the Faraday configuration \cite{Wang} and our calculation results for Cu$_2$O quantum well, obtained for $S$ exciton with $m=0$.  For effective evaluation of two very different systems, we use the dimensionless parameter $\gamma$ with appropriate Rydberg energy (87.78 meV for Cu$_2$O and 8.1 meV for GaAlAs\cite{Wang}). Furthermore, the significant difference in energy necessitates two y axes to overlap the data. This way, one can observe several similarities. In both systems, the magnetic field induced shift is quadratic in the low field regime and transitions to linear at $\gamma \sim 0.08$, which is consistent with our estimations. It should be stressed out that the results are accurate up to a constant; apart from the difference of band gaps and Rydberg energies, the data for GaAlAs is measured for quantum dots and our calculations have cylindrical symmetry. However, as pointed out in \cite{Wang} $^,$ \cite{Ziemkiewicz_PRB_2020}, proper adjustment of quantum well size allows for an approximation of a quantum dot, which is sufficient for the sake of presented comparison.

\begin{figure}[ht!]
\includegraphics[width=.9\linewidth]{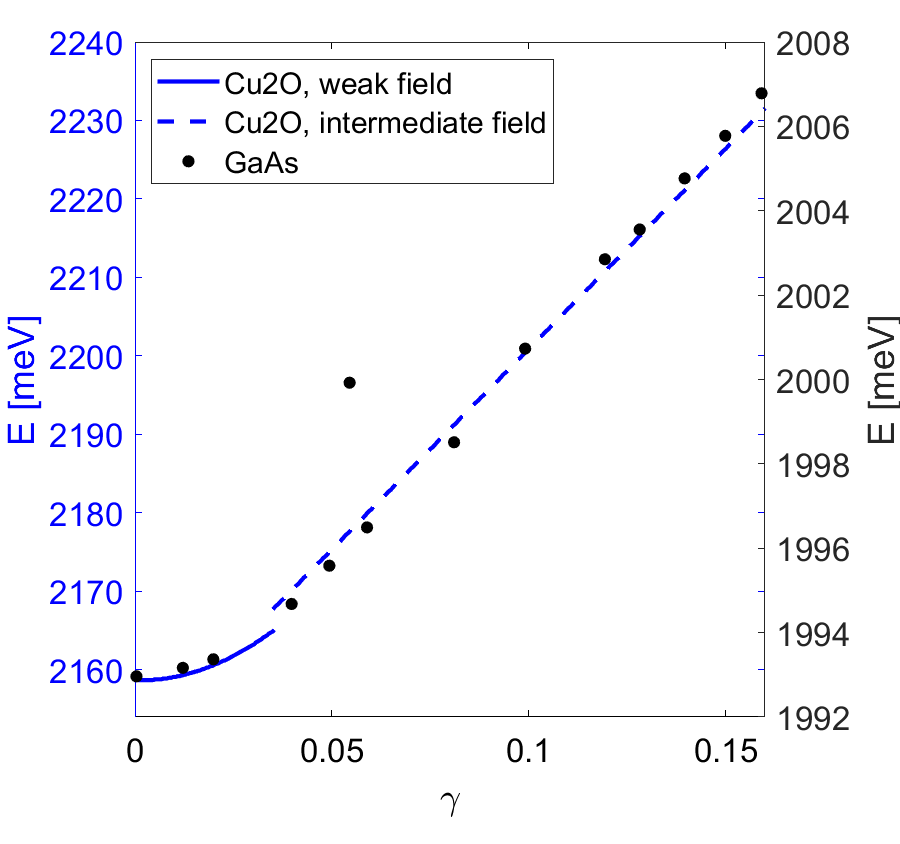}
\caption{Comparison of the calculated line shape and experimental results by Wang et al \cite{Wang} for GaAlAs.}
\label{fig6}
\end{figure}

The results for the Voigt configuration are compared with InAlAs on the Fig. \ref{fig7}. Again, the estimated boundary between weak and strong field regimes provides a good match to the experimental data.
\begin{figure}[ht!]
\includegraphics[width=.9\linewidth]{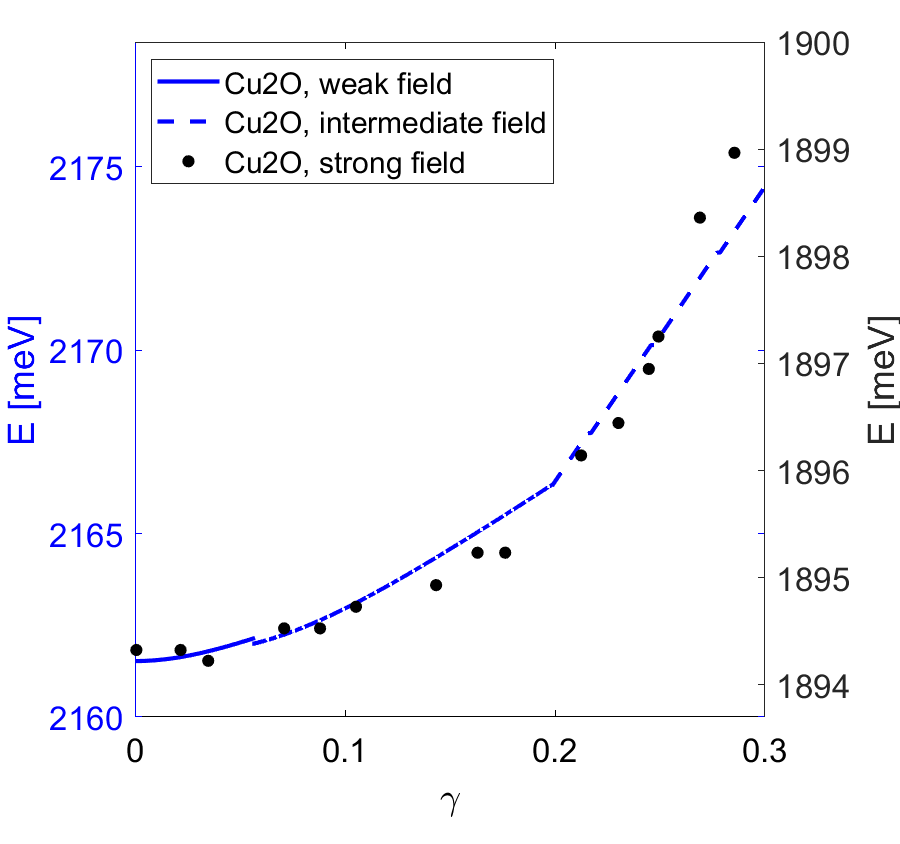}
\caption{Comparison of the calculated line shape and experimental results by Wang et al \cite{Wang} for InAlAs.}
\label{fig7}
\end{figure}

Finally, we use the experimental results of Jeon et al \cite{Jeon} to study the effect of well thickness on the energy, marked by $\Delta E=E(B)-E(0)$. One can observe an increase of the confinement energy with thickness $L$. For any fixed value of $B$, reduction of $L$ increases the energy (See Fig. \ref{fig5}); however, confinement states in larger quantum well exhibit stronger reaction to magnetic field, which results in higher energy overall.
\begin{figure}[ht!]
\includegraphics[width=.9\linewidth]{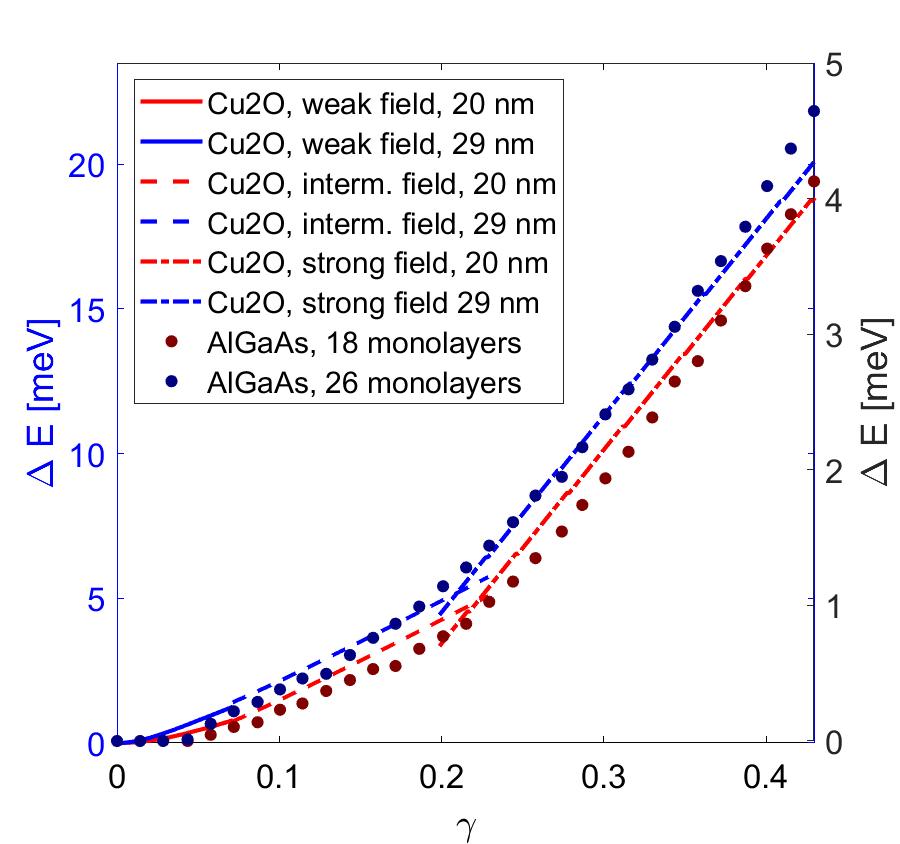}
\caption{Comparison of the calculated line shape and experimental results by Jeon et al \cite{Jeon} for GaAs/AlGaAs quantum well.}
\label{fig8}
\end{figure}

\section{Conclusions}\label{conclusions}
In the present work we have studied the magneto-optical functions for Cu$_2$O quantum wells with Rydberg excitons at two different orientations of the magnetic field.
A theoretical solutions to model absorption spectra due to excitons of Cu$_2$O in a quantum well in a wide range of magnetic fields is presented, with separate treatment of low, medium and high field regime. The theoretical analysis is done for both Faraday and Voigt external field configuration, including Landau splitting. We observe considerable inter-level mixing and splitting caused by differences in energy shifts of various excitonic states caused by confinement and magnetic field. Key characteristics of Cu$_2$O excitons - unusually high Rydberg energy, exceptionally large size of higher states and unique ratio of electron to hole mass all play a crucial role in forming rich magnetic absorption spectra. We conclude that in QW the difference between spectra obtained in both configurations depends on degrees of freedom involved in the interaction between the excitons and the magnetic field. Because of that   and  an effective electron and hole masses ratio we observe  that the boundaries of intermediate and strong field regime are significantly higher for the Voigt configuration.

Finally, we introduce a field-dependent parameter $\gamma$ which is a versatile tool for qualitative separation of the magnetic field regimes. Due to its universal nature, it can be employed to compare the calculation results with experimental spectra measured in other semiconductors, serving as a benchmark of the paper.

\section*{Acknowledgments} Support from National Science Centre,
Poland (project OPUS, CIREL  2017/25/B/ST3/00817) is greatly
acknowledged.
\appendix
\section{Matrix elements}\label{Appendix A}
We calculate the matrix elements (\ref{matrix_elem_Farad_week})
using the eigenfunctions (\ref{2_Dim_Eigen}). First we calculate
the diagonal elements

\begin{eqnarray}
&&\;V_{jj}=\frac{\gamma^2}{4}\langle R_{j1}(\rho)\vert \rho^2\vert
R_{j1}(\rho)\rangle\\
&&=\frac{\gamma^2}{4}16\lambda^3\left[\frac{j!}{(j+2\vert
m\vert)!}\right]\int\limits_0^\infty \rho\,d\rho
e^{-4\lambda\rho}(4\lambda\rho)^2\rho^2
\left[L_j^2(4\lambda\rho)\right]^2\nonumber\\
&&=\frac{\gamma^2}{64\lambda(j+1)(j+2)}\int\limits_0^\infty
dx\,e^{-x}x^4\left[xL^2_j(x)\right]L^2_j(x).\nonumber
\end{eqnarray}
The integral in the above equation is known\cite{Grad}, and we
obtained the expression
\begin{eqnarray}
&&V_{jj1}=\frac{\gamma^2(j+2)!}{64\lambda(j+1)(j+2)}4!\\
&&\times\left\{(2j+3)(j+2)!\left[\frac{1}{j!2!}\right]^2+\frac{(j+3)!}{[j!]^2\,3!}+
\frac{(j+1)(j+2)}{(j-1)!3!}\right\}.\nonumber
\end{eqnarray}
The off-diagonal elements can be obtained using Rodrigues$^,$
formula
\begin{equation}\label{Rodrig}
L_j^\alpha(x)=\sum\limits_{\ell=0}^j(-1)^\ell{j+\alpha\choose
j-\ell}\frac{x^\ell}{\ell!}.\end{equation} Performing the
integration we obtain the matrix elements in the form
\begin{eqnarray}\label{M_Elem_weak}
&&V_{ij1}=\frac{\gamma^2}{4}\int\limits_0^\infty \rho\,d\rho
R_{i1}(\rho)\,\rho^2\,R_{j1}(\rho)\nonumber\\
&&=4^3\gamma^2(\lambda_i\lambda_j)^{5/2}\left[\left(\frac{i!}{(i+2)!}\right)
\left(\frac{j!}{(j+2)!}\right)\right]^{1/2}\\
&&\times\sum\limits_{r=0}^i\sum\limits_{s=0}^j\Biggl\{{i+2\choose
i-r}{j+2\choose
j-s}\frac{(-1)^{r+s}}{r!s!}\nonumber\\
&&\times
(4\lambda_i)^r(4\lambda_j)^s\frac{(5+r+s)!}{[2(\lambda_i+\lambda_j)]^{5+r+s+1}}\Biggr\}.\nonumber
\end{eqnarray}
Using the above formula (\ref{Rodrig}) we calculated the matrix
elements $V_{n\ell m}$ for the high field limit (\ref{coeffHF}).
After simple transformations, for the case $\vert m\vert=1$, they
can be put into the form
\begin{displaymath}
V_{jk1}=-\frac{2\sqrt{2\gamma}}{\sqrt{(j+1)(k+1)}}\int\limits_0^\infty
dx\,e^{-x^2}x^2\,L^1_j(x^2)L_k^1(x^2),
\end{displaymath}
from which one obtains the formula

\begin{eqnarray}\label{matrix_elem_HF}
&&V_{jk1}=-\frac{1}{\sqrt{(j+1)(k+1)}}\sqrt{\frac{\pi\gamma}{2}}\\
&&\times\sum\limits_{r=0}^j\sum\limits_{s=0}^k{j+1\choose
j-r}{k+1\choose
k-s}\frac{(-1)^{r+s}}{r!s!}\frac{(2r+2s+1)!!}{2^{r+s}}.\nonumber
\end{eqnarray}
\section{Intermediate fields, Faraday configuration}\label{Appendix B}
Substituting the trial function (\ref{trial_Faraday}) into the Eq.
(\ref{Lippmann1}), with $V=-2/\rho$, one obtains the following
integral equation
\begin{eqnarray}\label{LS2}
&&\Psi_{00}R_{01}(\rho)\left[\sum\limits_{m=\pm
1}Y_{0m,00}\exp(-\kappa_{0m00}\rho)\frac{e^{im\phi}}{\sqrt{2\pi}}\right]\nonumber\\
&&+\sum\limits_{n=1}^\infty\sum\limits_{N_eN_h\geq 1}\sum\limits_m
\frac{e^{im\phi}}{\sqrt{2\pi}}Y_{nmN_eN_h}
R_{nm}(\rho)\Psi_{N_eN_h}\\
&&=\frac{2\mu}{\hbar^2
a^*}\left[(M_0\rho_0)\frac{2\gamma}{\sqrt{\pi}}\right]\frac{e^{i\phi}}{\sqrt{2\pi}}\sum\limits_{n=0}^\infty
R_{n1}(\rho)\frac{d_{n1}}{\kappa^2_{n1;N_eN_h}}\nonumber\\
&&+\frac{2\mu}{\hbar^2
a^*}\left[(M_0\rho_0)\frac{2\gamma}{\sqrt{\pi}}\right]\frac{e^{-i\phi}}{\sqrt{2\pi}}\sum\limits_{n=0}^\infty
R_{n1}(\rho)\frac{d_{n1}}{\kappa^2_{n,-1;N_eN_h}}\nonumber
\end{eqnarray}
\begin{eqnarray*}
&&+\int\limits_0^\infty \rho'
d\rho'\int\limits_0^{2\pi}d\phi'\int\limits_{-\infty}^\infty
dz_e'\int\limits_{-\infty}^\infty
dz_h'\Biggl\{G(\rho,\rho';\phi,\phi';z_e,z_e';z_h,z_h')\nonumber\\
&&\times\frac{2}{\rho'}R_{01}(\rho')\Psi_{00}(z_e',z_h')\\
&&\times\left[\sum\limits_{m=\pm
1}Y_{0m;00}\frac{\exp(-\kappa_{0m;00}\rho')}{\kappa^2_{nm;N_eN_h}}\frac{e^{im\phi'}}{\sqrt{2\pi}}\right]\Biggr\}.\nonumber
\end{eqnarray*}
From various methods of solving integral equations we choose the
method of projection on an orthonormal basis
$u_{nm}(\rho,\phi),\Psi_{N_eN_h}(z_e,z_h)$. We can use the
functions $\psi_{nm}(\rho)\exp(im\phi)/\sqrt{2\pi}$, to obtain
\begin{eqnarray}
&&Y_{0m,00}\langle R_{01}^2\vert
e^{-\kappa_{0m;00}\rho}=\frac{2\mu}{\hbar^2
a^*}\left[(M_0\rho_0)\frac{2\gamma}{\sqrt{\pi}}\right]{\mathcal
E}\frac{d_{01}}{\kappa^2_{0m;00}}\nonumber\\
&&+2\sum\limits_{m=\pm 1}Y_{0m,00}\int\limits_0^\infty
d\rho'\frac{e^{-\kappa_{0m;00}\rho'}}{\kappa^2_{0m;00}}R_{01}^2(\rho').\end{eqnarray}
From the above equation the parameters $Y_{0m,00}$ and $Y_{nm,00}$
(\ref{Y0Faraday}),(\ref{YnFaraday}) were obtained.
\section{The coefficients for the adiabatic potentials}\label{Appendix C}
The coefficients $a_{2n+1}$ are defined from the relations
\begin{eqnarray*}
&&\langle
\psi^{(1D)}_{\beta,N_y}(y)\vert\frac{2}{\sqrt{x^2+y^2}}\vert
\psi^{(1D)}_{\beta,N_y}(y)\rangle=\frac{2}{\vert x\vert
+a_{N_y}},\\
&&\frac{1}{a_{N_y}}=\left.\langle
\psi^{(1D)}_{\beta,N_y}(y)\vert\frac{2}{\sqrt{x^2+y^2}}\vert
\psi^{(1D)}_{\beta,N_y}(y)\rangle\right|_{x=0}.
\end{eqnarray*}
For odd parity eigenfunctions $N_y=2n+1$, and we use the relation
between Hermite polynomials and the confluent hypergeometric
function
\begin{eqnarray}
&&\psi^{(1D)}_{2n+1,\beta}(y)=A_{2n+1}H_{2n+1}(\beta
y)e^{-\beta^2y^2/2}\\
&&=A_{2n+1}(-1)^n\,2\frac{(2n+1)!}{n!}\beta
y\,M\left(-n,\frac{3}{2},\beta^2
y^2\right)e^{-\beta^2y^2/2}\nonumber\end{eqnarray} with the
normalization factor $A_{2n+1}.$ The coefficients $a_{2n+1}$ are
obtained from the following calculations
\begin{eqnarray}
&&\frac{1}{a_{2n+1}}=2\int\limits_0^\infty
\left[\psi^{(1D)}_{2n+1,\beta}\right]^2\frac{1}{y}dy\nonumber\\
&&=2\,[A_{2n+1}]^2\int\limits_0^\infty\left[\frac{(2n+1)!}{n!}\right]^24
\left[M\left(-n,\frac{3}{2},\beta^2y^2\right)\right]^2\nonumber\\
&&\times
e^{-\beta^2y^2}\beta^2 y\,dy\\
&&=\pi^{-1/2}\frac{\beta\,(2n+1)!}{2^{2n+1}(n!)^2}4\int\limits_0^\infty
e^{-z}\left[M\left(-n,\frac{3}{2},z\right)\right]^2\,dz\nonumber\\
&&=\pi^{-1/2}\frac{\beta\,(2n+1)!}{2^{2n-1}(n!)^2}J^n_1.\nonumber\end{eqnarray}
We use the integral\cite{Landau}
\begin{eqnarray*}
&&J^n_\nu=\int\limits_0^\infty e^{-kz}z^{\nu-1}[M(-n,\gamma,kz)]^2
dz,\nonumber\\
&&n=0,\\
&&J_\nu^0=\frac{1}{k^\nu}\Gamma(\nu),\nonumber\\
&&n=1,2,\ldots,\nonumber\\
&&J^n_\nu=\frac{\Gamma(\nu)n!}{k^\nu\gamma(\gamma+1)
\ldots(\gamma+n-1)}\left\{1+\frac{n(\gamma-\nu-1)(\gamma-\nu)}{1^2\cdot\gamma}
\right.\nonumber\\
&&\left.+\frac{n(n-1)(\gamma-\nu-2)(\gamma-\nu-1)(\gamma-\nu)(\gamma-\nu+1)}{1^2\cdot
2^2\cdot\gamma(\gamma+1)}+\ldots+\right.\nonumber\\
&&\left.+\frac{n(n-1)\ldots 1(\gamma-\nu-n)\ldots
(\gamma-\nu+n-1)}{1^2\ldots n^2\cdot \gamma(\gamma+1)\ldots
(\gamma+n-1)}\right\}\nonumber.\end{eqnarray*}

In our case we put $k=1,\quad \nu=1,\quad \gamma=\frac{3}{2}$. For
the lowest values of $n$ one obtains
\begin{eqnarray*}
&&n=0,\quad J_1^0=1,\quad \frac{1}{a_1}=\pi^{-1/2}\cdot 2\beta,\quad a_1=\frac{1}{2}\sqrt{\pi}\beta^{-1},\\
&&n=1,\qquad J_1^1=\frac{5}{9},\qquad
\frac{1}{a_3}=\pi^{-1/2}\cdot\frac{5\beta}{3},\quad a_3=\frac{3}{5}\sqrt{\pi}\beta^{-1},\\
&&n=2,\quad J_1^2=\frac{2}{5},\quad \frac{1}{a_5}=\pi^{-1/2}\cdot
\frac{3}{2}\beta,\quad a_5=\frac{2}{3}\sqrt{\pi}\beta^{-1}.
\end{eqnarray*}
\section{Determination of parameters, Voigt configuration,
intermediate fields}\label{Appendix_D} Inserting the trial
function (\ref{expansionV}) into Eq. (\ref{Lippmann2}), and using
the Green function (\ref{Green_V_intermediate}), one obtains the
equation and retaining the lowest expansion term in $GVY$ one
obtains the following expression
\begin{eqnarray*}\label{eqzetazero}
&&Y_{0}\Psi_{00}(z_e,z_h)\psi_{1,\beta}^{(1D)}(y)
e^{-\kappa_{0}\sqrt{x^2+y^2}}
\nonumber\\&&+\sum\limits_{n=1}^\infty \sum\limits_{N_eN_h\geq
1}\psi_{2n+1,\beta}^{(1D)}(y)\Psi_{N_eN_h}(z_e,z_h)\frac{1}{2\pi}\int\limits_{-\infty}^\infty
dk\,Y_{nN_eN_h}(k) e^{ikx}\\
&&=\frac{2\mu}{\hbar^2 a^*}{\mathcal
E}\frac{(M_0\rho_0)}{\sqrt{2\pi}}\sum\limits_n\sum\limits_{N_e}\sum\limits_{N_h}g_{2n+1}\psi_{2n+1,\beta}^{(1D)}(y)\\
&&\times\Psi_{N_eN_h}(z_e,z_h)\delta_{N_eN_h}\int\limits_{-\infty}^\infty
\frac{dk e^{ikx}\,e^{-k^2\rho_0^2/2}}{k^2+\kappa_{nN_eN_h}^2}\\
&&+4Y_0\sum\limits_{n,N_e,N_h}\int\limits_{-\infty}^\infty
dz_e'\int\limits_{-\infty}^\infty
dz_h'\\
&&\times\frac{1}{2\pi}\int\limits_{0}^\infty
dx'\int\limits_{-\infty}^\infty dy'\int\limits_{-\infty}^\infty
dk\,e^{ikx}\cos
kx'\,\psi^{(1D)}_{\beta,n}(y)\psi^{(1D)}_{\beta,n}(y')\\
&&\times\frac{\psi^{(1D)}_{\alpha_e^V,N_e}(z_e)\psi^{(1D)}_{\alpha_e^V,N_e}(z_e')
\psi^{(1D)}_{\alpha_h^V,N_h}(z_h)\psi^{(1D)}_{\alpha_h^V,N_h}(z'_h)}{k^2+\kappa_{nN_eN_h}^2}\\
&&\times
\frac{\exp(-\kappa_0\sqrt{x'^2+y'^2})}{\sqrt{x'^2+y'^2}}\psi^{(1D)}_{\beta,1}(y')\Psi_{00}(z_e'z_h')\\
\end{eqnarray*}
Similar equation has been obtained in Appendix \ref{Appendix B},
and was solved by making projections on a orthonormal set of
functions. Here we choose the functions
$\{\psi^{(1D)}_{\beta,n}(y)\}, \Psi_{N_eN_h}$, put $x=0$,  and
obtain
\begin{eqnarray}\label{Fkappa}
&&Y_0=\frac{2\mu}{\hbar^2 a^*}{\mathcal
E}\frac{(M_0\rho_0)}{\sqrt{\pi}}\frac{ {g_{1}}}{\kappa_0}\nonumber\\
&&\times\left[\frac{2^{3/2}}{\sqrt{\pi}}e^{\kappa_0^2/8\beta^2}D_{-3}\left(\frac{\kappa_0}{\beta\sqrt{2}}\right)
-F(\kappa_0,\beta)\right]^{-1},\\
&&F(\kappa_0,\beta)=\frac{\beta}{\sqrt{\pi}}\int\limits_{-\infty}^\infty\,dk\,
 \frac{e^{-k^2\rho_0^2/2}}{(k^2+\kappa_0^2)^{3/2}}\nonumber\\
 &&\times
 \exp\left(\frac{k^2+\kappa_0^2}{8\beta^2}\right)W_{-1,0}\left(\frac{k^2+\kappa_0^2}{4\beta^2}\right),\nonumber\\
 &&\frac{1}{2\pi}Y_{nN}(k)=\frac{2\mu}{\hbar^2 a^*}{\mathcal
 E}\frac{M_0\rho_0}{\sqrt{2\pi}}g_{2n+1}\,\frac{e^{-k^2\rho_0^2}}{k^2+\kappa_{nN}^2}.\nonumber
\end{eqnarray}
The above quantities, substituted in Eq. (\ref{expansionV}),
determine the exciton amplitude $Y$, which inserted in Eq.
(\ref{def_mean_susceptibility}), gives the magneto-susceptibility
(\ref{chiVintermediate}).

\end{document}